\documentclass[journal=jctcce,manuscript=article]{achemso}

\usepackage[version=3]{mhchem} 
\usepackage[T1]{fontenc}
\usepackage{tipa}
\usepackage{hyperref}

\usepackage{amssymb, amsmath}
\usepackage{braket}
\usepackage{bm}
\usepackage{upgreek}
\newcommand{\mr}{\mathrm}

\newcommand*{\matr}[1]{\mathbf{#1}}
\newcommand*{\vect}[1]{\bm{#1}}

\usepackage{algorithm}
\usepackage{algpseudocode}

\usepackage[authormarkuptext=name,authormarkup=none,final]{changes}
\definecolor{red}{RGB}{255, 0, 0}
\definecolor{blue}{RGB}{0, 0, 255}
\definecolor{green}{RGB}{0, 192, 0}
\definechangesauthor[name={Authors}, color={blue}]{GL}
\definechangesauthor[name={Authors}, color={blue}]{YLAS}

\usepackage{tablefootnote}
\usepackage{fancyhdr}

  \newcommand*{\citen}{}
\DeclareRobustCommand*{\citen}[1]{%
  \begingroup
    \romannumeral-`\x 
    \setcitestyle{numbers}%
    \cite{#1}%
  \endgroup
}

\author{Yorick L. A. Schmerwitz}
\affiliation{Max-Planck-Institut f\"ur Kohlenforschung, 45470 M\"ulheim an der Ruhr, Germany}
\email{schmerwitz@kofo.mpg.de}
\author{Elli Selenius}
\affiliation{Science Institute of the University of Iceland, 107 Reykjavík, Iceland}
\author{Gianluca Levi}
\affiliation{Department of Chemical and Pharmaceutical Sciences, University of Trieste, 34127 Trieste, Italy}
\alsoaffiliation{Science Institute of the University of Iceland, 107 Reykjavík, Iceland}
\email{gianluca.levi@units.it}

\title{Freeze-and-release direct optimization method for variational calculations of excited electronic states}

\begin{document}

\begin{tocentry}
   \includegraphics[width=\textwidth]{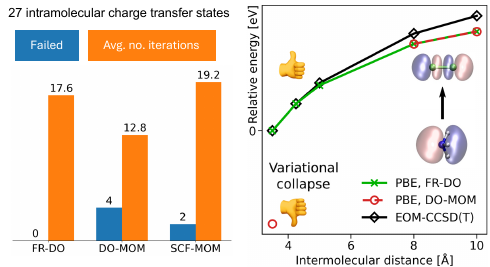}
\end{tocentry}

\begin{abstract}
Variational optimization of orbitals in time-independent density functional calculations of excited electronic states presents a significant challenge, as excited states typically correspond to saddle points on the electronic energy landscape. The optimization can be particularly difficult if the excitation involves significant rearrangement of the electron density, as for charge transfer excitations. A simple strategy for variational orbital optimization of excited states is presented. The approach involves minimizing the energy while freezing the orbitals directly involved in the excitation, followed by a fully unconstrained saddle point optimization. Both steps of this freeze-and-release strategy are carried out using direct optimization algorithms with the same computational scaling as ground state calculations. The performance of the method is extensively assessed in calculations of intramolecular and intermolecular charge transfer excited states of organic molecules and molecular dimers using a generalized gradient approximation functional. It is found that the freeze-and-release direct optimization approach can avoid variational collapse to spurious, charge-delocalized solutions for cases where conventional algorithms based on the maximum overlap method fail. For intermolecular charge transfer, the orbital-optimized calculations are found to provide the correct dependency of the energy on the donor-acceptor separation without requiring long-range exact exchange, something common time-dependent density functional theory approaches fail to achieve.
\end{abstract}


\section{Introduction}
Accurate modeling of electronic excitations in large molecular chromophores and photoactive materials is essential for understanding phenomena such as photosynthesis and vision, as well as for developing efficient and sustainable solar energy conversion devices. Photoinduced processes typically involve multiple states with different electronic character, with charge transfer excited states often playing a key role. Thus, simulating photoinduced processes relevant to both natural phenomena and solar energy devices requires electronic structure methods that are both computationally affordable and can reliably describe states of different electronic character across different molecular geometries. 
While numerous wave-function-based and density functional theory (DFT) approaches for excited state calculations exist, achieving the required accuracy, and efficiency for routine application remains a fundamental challenge. This is particularly the case for charge transfer excitations\cite{Knepp2025}, which are characterized by a large rearrangement of the electron density with respect to that of the ground state. As a result, conventional methods typically hold limited predictive power in simulations of the photoinduced dynamics of electrons and nuclei\cite{Mukherjee2024, Janos2023}, calling for continued development of efficient excited state methodologies with sufficient accuracy.

The most favorable balance between efficiency and accuracy is typically achieved in density-functional-based calculations. These calculations are most efficient when employing local or semilocal Kohn-Sham (KS)\cite{Kohn1965, Hohenberg1964} functionals, which approximate the electron-electron exchange and correlation through density-based exchange-correlation (XC) functionals. However, recent algorithmic advances have reduced the computational cost of more elaborate hybrid functionals as well, which incorporate a portion of non-local exact exchange explicitly\cite{Helmich-Paris2021, DAntoni2023, Rebolini2016}. 

Within the density functional framework, excited states are most frequently modelled using time-dependent DFT (TDDFT)\cite{Runge1984, Hohenberg1964}. Practical implementations of TDDFT are usually based on linear-response perturbation theory applied to the time-dependent KS equations\cite{Casida1995}, using ground state functionals  within the adiabatic approximation\cite{Maitra2016}. While comparable in efficiency to the time-independent ground state DFT formalism, the approximations used in TDDFT significantly limit its applicability in calculations of excited states involving large rearrangement of the electron density, including charge transfer\cite{Mester2022, Dev2012, Ghosh2015, Dreuw2004} and Rydberg\cite{Cheng2008} excitations. In particular, TDDFT tends to underestimate the excitation energy of charge transfer excitation and fails to provide the correct, $1/R$, variation of the energy as a function of the separation $R$ between donor and acceptor fragments\cite{Mester2022, Dev2012, Ghosh2015, Dreuw2004}. These failures are rooted in the lack of derivative discontinuity in standard KS functionals as well as the lack of divergence of the adiabatic XC kernel as $R\rightarrow\infty$\cite{Maitra2022, Hellgren2012}. Nonadiabatic approximations offer a promising route to address these flaws of TDDFT, but are still in the early stages of development\cite{Lacombe2023}. Range-separated hybrid functionals with optimally tuned range-separation parameters\cite{Stein2009}, which recover full exact exchange at long range, can also improve the description of charge-transfer excitations in TDDFT \cite{Bogo2024}. The challenge there, however, is that the optimal tuning is system-specific, may not generalize across different states, and can introduce discontinuities in potential energy surfaces.\cite{Bogo2025, Maitra2022} Indeed, a generally applicable TDDFT approach for treating charge transfer excitations is not yet available.

The limitations of practical TDDFT implementations have spurred the development of time-independent DFT formalisms for excited states within both ensemble\cite{Cernatic2024, Gould2021, Cernatic2021, Gould2018, Yang2017, Theophilou1979} and state-specific\cite{Ayers2015, Ayers2012, Ayers2009, Levy1999, Gorling1999, Perdew1985} frameworks. Recently, the first state-specific\cite{Loos2025, Gould2025} and ensemble\cite{Gould2023} density functionals designed for variationally optimized excited states have been developed, representing an important step forward in excited state DFTs. Despite these theoretical advancements, time-independent excited state density functional calculations are typically based on a practical orbital-optimized (OO) approach\cite{Hait2021}, sometimes referred to as $\Delta$Self-Consistent Field ($\Delta$SCF), where excited states are found as nonaufbau solutions to the time-independent KS equations beyond the lowest energy (ground state) solution using ground state XC functionals. OO-KS calculations of excited states find partial justification in adiabatic TDDFT, where nonaufbau KS solutions correspond to stationary densities\cite{Kowalczyk2011}. More recently, Gould\cite{Gould2025pra} and Fromager\cite{Fromager2025} have shown that the approach can be derived from an exact stationarity condition for ground and excited states with respect to the ensemble density when the ensemble XC functional is constructed from the regular ground state functional, providing further theoretical underpinning. Yang and Ayers\cite{Yang2024} have also recently provided a rationale based on functionals that depend on the KS potential.

While these recent developments provide a theoretical foundation, the growing interest in OO density functional methods for excited states has largely been driven by their practical success. Compared to TDDFT, these methods usually strike a more balanced description of ground and excited states as they are treated on an equivalent variational footing. In particular, OO-KS calculations outperform TDDFT for challenging excitations, such as doubly and core-level excited states\cite{Hait2021, Hait2020}, charge transfer\cite{Bogo2025, Froitzheim2024, Bogo2024, Selenius2024, Barca2018} and Rydberg excitations\cite{Sigurdarson2023, Cheng2008}, and there are preliminary indications that they can better describe conical intersections between ground and excited states\cite{Schmerwitz2022, Barca2018}. Another advantage of OO excited state approaches compared to TDDFT is that tasks such as the computation of atomic forces or the inclusion of solvation effects through both implicit\cite{Froitzheim2024} and explicit models\cite{Mazzeo2023, Vandaele2022, Levi2020pccp, Levi2018} can be done using techniques developed and implemented for ground state methods. 

Typically, KS excited state solutions correspond to saddle points on the electronic energy surface\cite{Schmerwitz2023}. As a result, OO-KS calculations are susceptible to variational collapse along directions of negative curvature, leading to convergence to solutions lower in energy than the target excited state. This issue remains one of the most significant practical challenges in the application of the methodology. The most widely used approach to mitigate the variational collapse is the maximum overlap method (MOM)\cite{Barca2018, Gilbert2008, Cheng2008}, which attempts to preserve a given nonaufbau configuration by occupying, at each wave function optimization step, the orbitals that overlap the most with a set of reference orbitals, typically chosen as the initial guess orbitals\cite{Barca2018}. \replaced[id=GL, comment=R1C2]{Various MOM approaches have been developed and employed in combination with both SCF algorithms based on eigendecomposition of the Hamiltonian matrix\cite{Corzo2022, Daga2021, Macetti2021, Obermeyer2021, Barca2018, Gilbert2008}, such as the direct inversion in the iterative subspace (DIIS)\cite{Pulay1980}, and direct optimization approaches, which directly seek a unitary transformation yielding the optimal orbitals of the target stationary solution\cite{Kumar2022, Ivanov2021, Levi2020a, Levi2020b}.}{MOM has been employed in combination with both SCF algorithms based on eigendecomposition of the Hamiltonian matrix\cite{Sinyavskiy2025, Corzo2022, Obermeyer2021, Barca2018, Gilbert2008}, such as the direct inversion in the iterative subspace (DIIS)\cite{Pulay1980}, and direct optimization approaches, which directly seek a unitary transformation yielding the optimal orbitals of the target stationary solution\cite{Ivanov2021, Levi2020a, Levi2020b}.} Another commonly used strategy involves the application of \replaced[id=GL, comment=R1C2]{a level shifting\cite{Kumar2022, Carter2020} within DIIS}{a level shifting\cite{Carter2020} within DIIS}. Level shifting raises the energy of the unoccupied orbitals, thereby restricting the occupied-unoccupied orbital rotations and reducing the risk of variational collapse. Another strategy consists in recasting the saddle point optimization as a minimization by minimizing the square of the gradient instead of the gradient itself\cite{Hait2020}. This and related approaches based on minimizing the Hamiltonian variance\cite{Zhao2019, Ye2017} face the challenge that minima on the variance optimization landscape are connected by unphysical stationary points with small energy barriers, which can lead to instabilities and convergence to incorrect solutions\cite{Bogo2024, Burton2022, Cuzzocrea2020}. Moreover, the computational cost is increased compared to ground state calculations due to the need of computing the gradient of the squared energy gradient. Finally, constrained methods have been developed that enforce the orthogonality of the target excited state with all lower energy states\cite{Pham2025, Evangelista2013}. While these methods prevent variational collapse by construction, they require the calculation of all states up to the desired one, making them best suited for the lowest excited states.

While conventional methods, such as MOM, are easy to implement and reduce the risk of variational collapse, they do not eliminate it entirely and can lead to convergence problems, especially when used in combination with DIIS algorithms\cite{Bogo2024, Schmerwitz2023, Mazzeo2023, Ivanov2021, Carter2020, Hait2020, Levi2020b, Mewes2014}. Issues are observed particularly for excited states with charge transfer character, including both intramolecular\cite{Selenius2024, Schmerwitz2023, Mazzeo2023, Ivanov2021, Carter2020, Hait2020, Levi2020b, Mewes2014} and intermolecular\cite{Bogo2025, Bogo2024} charge transfer excitations. In such cases, OO-KS calculations often exhibit oscillatory convergence behavior or tend to collapse to lower-energy solutions where the charge is delocalized and the charge transfer character is therefore less pronounced\cite{Bogo2025, Bogo2024, Selenius2024}. This represents a significant limitation for the currently most used strategies for OO-KS excited state calculations, since charge transfer excitations play a fundamental role in both biological processes and applications for solar energy conversion. Nevertheless, OO-KS calculations have been proposed for high-throughput screenings of photofunctional molecules and materials\cite{Bogo2024, Froitzheim2024}, as they do not rely on sophisticated and computationally demanding density functional approximations to provide reasonably accurate results for charge transfer excitations (when convergence is not an issue). This underscores the need for improved excited state orbital optimization strategies that are robust and also straightforward to implement and apply in practice.

Recently, we have introduced\cite{Selenius2024, Schmerwitz2023} a simple two-step strategy for OO calculations of excited states involving (1) a minimization of the energy along all degrees of freedom except those along which the energy should be maximized, which are kept fixed, followed by (2) a fully unconstrained optimization designed to find a saddle point of the energy surface. The saddle point optimization in step (2) was carried out using either a quasi-Newton algorithm that can handle Hessians with negative eigenvalues\cite{Selenius2024}, or a generalized mode following (GMF) approach\cite{Schmerwitz2023} where the projection of the gradient along the modes of the electronic Hessian corresponding to the $n$ lowest eigenvalues is inverted in order to converge on a saddle point of order $n$. In both cases, the orbital optimization is performed via direct optimization with the exponential transformation\cite{Ivanov2021, Ivanov2021cpc, Levi2020b}. Here, this approach is referred to as freeze-and-release direct optimization (FR-DO), adopting a terminology introduced by Obermeyer et al.\cite{Obermeyer2021} for a two-step wave function optimization that applies constraints in the initial step. In our earlier works\cite{Selenius2024, Schmerwitz2023}, this FR-DO approach was used to asses the performance of the local density approximation (LDA) and generalized gradient approximation (GGA) functionals in orbital-optimized calculations of charge transfer excitations in organic molecules. More recently, Bogo et al.\cite{Bogo2025}, inspired by our work, have adopted a similar freeze-and-release scheme using geometric direct minimization\cite{Voorhis2002} for the constrained optimization step and squared-gradient minimization\cite{Hait2020} for the second step of fully unconstrained optimization to calculate long-range charge transfer states in molecular dimers as well as large supramolecular and dye-semiconductor complexes.

Although the previous studies indicate that the FR-DO approach can converge charge transfer excited states, a systematic assessment of its performance in comparison to conventional OO algorithms and rationalization of its success is lacking. Here, an extensive assessment of the convergence properties of the FR-DO strategy using a quasi-Newton algorithm for the saddle point search step is presented, and insights are provided into the factors governing its efficiency compared to previous approaches. The assessment is done through GGA calculations on a large set of charge transfer excited states of organic molecules\cite{Loos2021} as well as intermolecular charge transfer excitations in molecular dimers, which have not been investigated in previous FR-DO works. The conventional MOM strategy even when starting from localized orbitals, is demonstrated prone to a systematic failure, which leads to convergence to spurious solutions where the hole and excited electron orbitals are delocalized, significantly reducing the charge transfer character of the excitation. Moreover, when used in combination with DIIS, the calculations can exhibit erratic convergence. In contrast, FR-DO avoids the variational collapse. The improvement arises from the partial relaxation of the orbitals in the constrained optimization step, which provides a higher-quality initial guess from which the directions of negative curvature can be identified. While FR-DO typically requires more iterations than DO calculations with MOM, the same computational scaling as for ground state calculations is retained. Using the obtained FR-DO solutions, it is shown that OO calculations with a GGA functional predict the correct $1/R$ dependency of the excitation energy on the donor-acceptor separation for the intermolecular charge transfer states, something linear-response TDDFT fails to achieve.

The article is organized as follows. Section 2 presents a comprehensive description of the FR-DO method, along with the computational settings employed in the calculations. Section 3.1 presents the results of the calculations on intramolecular charge transfer excitations in organic molecules, comparing the convergence properties of FR-DO and conventional MOM approaches. The saddle point order of the target solutions is estimated before and after constrained optimization, and a representative case, the A$_1$ state of twisted $N$-Phenylpyrrole, is used to examine the failure of MOM and explain the improved performance of the FR-DO strategy. Section 3.2 presents the results of the calculations on intermolecular charge transfer states in molecular dimers. The main findings are summarized in section 4.

\section{Methodology}
\subsection{Direct orbital optimization}
In variational electronic structure methods, excited states are obtained as stationary points on the electronic energy surface\cite{Burton2022, MolecularElectronicStructureTheory}, \added[id=GL]{typically}\footnote{While the exact excited states provided by full configuration interaction always correspond to saddle points, with a saddle point order given by the level of the excitation, approximate wave functions have fewer degrees of freedom and may in some cases become local minima. A representative example is the symmetry-preserving doubly excited state solution of the H$_2$ molecule in a spin-restricted calculation with minimal basis set\cite{Schmerwitz2023}.} \added[id=GL, comment={R1C8, R2C1}]{saddle points \cite{Schmerwitz2023, Marie2023, Burton2022, Kossoski2021, Perdew1985}} as a variation in the electronic degrees of freedom driving the system toward the ground state (the global minimum) leads to a decrease in energy. In OO-KS calculations of excited states\cite{Hait2021}, the objective is to find a single-Slater-determinant wave function with a nonaufbau orbital occupation for which the energy functional of the electron density is stationary with respect to unitary variations of the orbitals. A common approach involves solving the KS equations for high-energy solutions via eigendecomposition of the KS Hamiltonian matrix, an extension of conventional SCF methods for ground state calculations. However, this approach often suffers from convergence failures and can struggle to maintain the desired nonaufbau occupation during the SCF, even when techniques such as MOM are employed\cite{Bogo2024, Schmerwitz2023, Mazzeo2023, Ivanov2021, Carter2020, Hait2020, Levi2020b, Mewes2014}, as also shown in the present work.

An alternative approach relies on direct orbital optimization (DO), where the orbitals are directly optimized by finding a unitary transformation that makes the KS energy stationary\cite{Ivanov2021, Ivanov2021cpc, Levi2020b, Lehtola2020, Voorhis2002, Head-Gordon1988}. Given a reference set of $M$ orthonormal molecular orbitals, $\vect{\psi}_{0} = \left\{\psi^0_i(\vect{r})\ |\ 1\leq i\leq M\right\}$, usually chosen as the initial guess orbitals and including both occupied and unoccupied orbitals, a new set of orthonormal orbitals $\vect{\psi} = \left\{\psi_i(\vect{r})\ |\ 1\leq i\leq M\right\}$ is obtained via the unitary transformation
\begin{equation}\label{eq:unitary}
 \vect{\psi} = \vect{\psi}_0\matr{U}\,.
\end{equation}
The unitary matrix $\matr{U}$ is commonly parametrized as the exponential of an anti-Hermitian matrix $\boldsymbol{\kappa} = - \boldsymbol{\kappa}^{\dagger}$, such that $\vect{\psi} = \vect{\psi}_0 e^{\boldsymbol{\kappa}}$. Therefore, in general, finding the optimal orbitals corresponding to an excited state solution involves making the energy stationary with respect to the elements of $\boldsymbol{\kappa}$ and simultaneously minimizing it with respect to $\vect{\psi}_0$, which leads to the variational condition\cite{Ivanov2021}
\begin{equation}
 \underset{\vect{\psi}} {\mathrm{stat}\,} E[\vect{\psi}] = \underset{\vect{\psi}_0} {\mathrm{min}\,}
 \underset{\boldsymbol{\kappa}} {\mathrm{stat}\,} E[\vect{\psi}_0e^{\boldsymbol{\kappa}}]\,.
\end{equation}

In the present work, the molecular orbitals are represented using a linear combination of atomic orbitals (LCAO) basis set, where the initial orbitals are expressed as a linear combination of $M$ basis functions
\begin{equation}
 \vect{\psi}_0 =\vect{\chi} \matr{C}_0\,,
\end{equation}
where $\vect{\chi} = \left\{\chi_i(\vect{r})\ |\ 1\leq i\leq M\right\}$ is the vector of basis functions, and $\matr{C}_0$ is the $M \times M$ matrix of expansion coefficients. Since the basis functions are fixed, the variational condition becomes
\begin{equation}
 \underset{\vect{\psi}} {\mathrm{stat}\,} E[\vect{\psi}] = \underset{\boldsymbol{\kappa}} {\mathrm{stat}\,} E[\vect{\chi} \matr{C}_0e^{\boldsymbol{\kappa}} ]
\end{equation}
Thus, in the LCAO basis set, the variational optimization of the orbitals reduces to finding the elements of $\boldsymbol{\kappa}$ that make the energy stationary, providing a matrix of optimal coefficients. 

Anti-Hermitian matrices $\boldsymbol{\kappa}$ form a linear space, which makes it possible to carry out the optimization with gradient-based optimization algorithms, as long as the unitary transformation in eq\ \ref{eq:unitary} is applied to the orbitals at every optimization step before the gradient is calculated\cite{Voorhis2002}. Efficient quasi-Newton methods that propagate a non-positive-definite approximate electronic Hessian have been proposed\cite{Levi2020b, Levi2020a}, and are currently in use\cite{Selenius2024, Sigurdarson2023, Ivanov2023}. The electronic gradient elements, $\partial E/\partial\kappa_{ij}$, are computed using the elements of the Hamiltonian (Fock) matrix in the basis of the optimal orbitals
\begin{align}
& H_{ij} = \sum_{\mu\nu} C^*_{i \mu} H_{\mu\nu} C_{\nu j}
\end{align}
with
\begin{equation}
H_{\mu\nu} = \int \chi_\mu^*(\vect{r}) {\bf \hat{h}}_{\mr{KS}} \chi_\nu(\vect{r}) \mathrm{d} \vect{r}\,,
\end{equation}
where ${\bf \hat{h}}_{\mr{KS}}$ is the KS Hamiltonian operator. The gradients evaluated at each step in the optimization are then used to propagate an approximate inverse electronic Hessian starting from an initial inverse Hessian that acts as a preconditioner for the quasi-Newton step. A common choice for the initial Hessian is a diagonal approximation with elements\cite{Ivanov2021cpc, Levi2020b, Head-Gordon1988}
\begin{equation}\label{eq:precond}
    \centering
    \frac{\partial^{2}E}{\partial \kappa_{ij}^{2}} \approx 2 \left( \epsilon_i- \epsilon_j \right) \left( f_j - f_i\right)\,,
\end{equation}
which is easily computed from the eigenvalues, $\epsilon_i$, and occupation numbers, $f_i$, of the canonical orbitals of the initial guess. Further details on the matrix exponential, gradient evaluation, and propagation of the approximate Hessian can be found in previous works\cite{Ivanov2021, Ivanov2021cpc, Levi2020b}.

Recently, an alternative direct optimization approach for excited states based on generalized mode following has been presented\cite{Schmerwitz2023}. This method involves determining the eigenvectors of the electronic Hessian corresponding to its lowest $n$ eigenvalues using a numeric partial diagonalization strategy. The components of the electronic gradient along the $n$ modes are then inverted, and a step uphill in energy in the directions parallel to the eigenvectors can be taken using established minimization methods, thereby converging on a saddle point of order $n$. This DO-GMF method is more robust than DO methods based on the quasi-Newton step, but requires an accurate estimate of the saddle point order of the target excited state solution.

\subsubsection{Direct optimization with constraints}
The direct optimization strategy illustrated above can readily be adapted to perform a constrained optimization where a subset of $N$ orbitals is relaxed while the remaining $M-N$ orbitals are kept fixed. Let $\vect{s} =\ \left\{s_i\ |\ 1\leq i\leq N\right\}$ be the set of indices of the $N$ orbitals that are optimized. Then, the elements of the matrix of coefficients after application of the unitary transformation are given by
\begin{align}
C_{\mu i} =
\begin{cases}
    \sum_{k=1}^N C^0_{\mu k}\left[ e^{\boldsymbol{\kappa}}\right]_{ki} \quad
    &\mr{for} \quad i \in \vect{s}
    \\
    C_{\mu i}^0 \quad
    &\mr{for} \quad i \notin \vect{s}
\end{cases}\,
\end{align}
During the constrained optimization, the gradient can be evaluated using the elements of a reduced $N \times N$ Hamiltonian matrix
\begin{equation}
H^\prime_{kl} = \sum_{\mu\nu} C^{\prime*}_{k \mu} H_{\mu\nu} C^\prime_{\nu l}\,,
\end{equation}
where $1\leq k,\ l\leq N$ are orbital indices in the subspace containing the relaxed orbitals.

\subsection{Freeze-and-release direct optimization}\label{sec:freeze_and_release}
The freeze-and-release direct optimization (FR-DO) method is summarized in Algorithm \ref{alg:fd-do}. 
\begin{algorithm}[h!]
\small
\caption{Freeze-and-Release Direct Optimization}\label{alg:fd-do}
\begin{algorithmic}[1]
\State\textbf{Input:}
\Statex \hspace{0.5cm} - Ground state orbitals $\{{\psi}^0_i\ |\ 1\leq i\leq M\}$
\Statex \hspace{0.5cm} - $P$ pairs of occupied–unoccupied orbital indices of target excitations $\left\{\left\{r_{p},\ a_{p}\right\}\ |\ 1\leq p\leq P\right\}$
\State \textbf{Initial guess excitations}
\Statex For all $1\leq p\leq P$, apply 90$^\circ$ rotation to swap occupation numbers of ${\psi}^0_{r_p}$ and ${\psi}^0_{a_p}$
\State \textbf{Constrained subspace optimization}
\Statex Freeze all hole orbitals $\vect{\psi}_{\mr{h}}^0 = \{\psi^0_{r_p}\ |\ 1\leq p\leq P\}$ and excited electron orbitals $\vect{\psi}_{\mr{e}}^0 = \{\psi^0_{a_p}\ |\ 1\leq p\leq P\}$ and optimize remaining $M-2P$ orbitals
\Statex $\hspace{11pt}\vect{\psi}^{\prime}\leftarrow\vect{\psi}^\prime e^{\boldsymbol{\kappa^\prime}}\hspace{19.5pt}\left\{{\psi}^\prime_k\ |\ 1\leq k\leq M - 2P\right\} = \left\{{\psi}^0_i\ |\ 1\leq i\leq M\right\} \setminus\left(\vect{\psi}_{\mr{h}}^0\cup\vect{\psi}_{\mr{e}}^0\right)\hfill$
\State \textbf{Electronic Hessian analysis}
\Statex Compute approximate Hessian in full space $\left\{{\psi}^\prime_k\ |\ 1\leq k\leq M - 2P\right\}\cup\left(\vect{\psi}_{\mr{h}}^0\cup\vect{\psi}_{\mr{e}}^0\right)$
\Statex Estimate directions of negative curvature and saddle point order of target excited state
\State \textbf{Full-space unconstrained optimization with saddle point search}  
\Statex Optimize all $M$ orbitals
\Statex $\quad \vect{\psi} \leftarrow \vect{\psi}e^{\boldsymbol{\kappa}}$
\State \textbf{Output:} Fully optimized excited state orbitals $\left\{{\psi}_i\ |\ 1\leq i\leq M\right\}$
\end{algorithmic}
\end{algorithm}
As commonly done, the initial guess for the excited state calculation is formed from ground state orbitals with occupation numbers modified to reflect a desired excitation. For instance, for a HOMO-LUMO excitation, a 90$^\circ$ rotation between the ground state HOMO and LUMO orbitals is performed, effectively swapping their occupation numbers. In general, the occupation numbers can be swapped between $P$ pairs of ground state occupied-unoccupied orbitals, corresponding to $P$ excitations. Next, a constrained optimization is performed where the orbitals involved in the excitations are kept fixed, constraining all pairwise rotations involving these orbitals. This step prevents variational collapse to lower-energy solutions while allowing the remaining orbitals to relax. As will be shown later, the partial relaxation makes it possible to produce an improved estimate of the directions of negative curvature that must be followed to locate the target excited state saddle point. Using the orbitals obtained from the constrained optimization as initial guess, along with the refined approximate electronic Hessian, a subsequent unconstrained optimization is performed in the full orbital space to converge on the saddle point of the target excited state.

This optimization strategy has some similarities with the freeze-and-release method presented by Obermeyer et al.\cite{Obermeyer2021} in the context of Hartree-Fock-Slater calculations of multiply ionized and highly excited states in molecules. Both approaches involve steps of constrained optimization where some orbitals are frozen, but there are some differences. Firstly, the FR-DO approach uses direct orbital optimization with quasi-Newton algorithms designed to converge on saddle points. The strategy by Obermeyer et al.\cite{Obermeyer2021}, instead, is based on solving the SCF eigenvalue equation in combinations with MOM. Secondly, in the strategy presented in ref \citen{Obermeyer2021}, the orbitals kept fixed in the constrained optimization steps are chosen based on the magnitude of the components of the energy gradient. Here, instead, a more bespoke definition of the constraints is adopted, as the fixed orbitals are chosen as the reference orbitals with holes and excited electrons based on the excitations with respect to the ground state orbitals. In this way, the constrained degrees of freedom contain the hard-to-predict degrees of freedom along which the energy should be maximized, allowing for a partial orbital relaxation without the risk of variational collapse.

The FR-DO method is implemented in the Grid-based Projector Augmented Wave (GPAW) software\cite{Mortensen2024, Enkovaara2010, Mortensen2005}.

\subsection{Computational settings}
Calculations with the FR-DO method as well as OO methods using MOM are carried out for intramolecular and intermolecular charge transfer states. The intramolecular charge transfer excitations consist of a set of 27 states of 15 organic molecules, as identified in ref \citen{Loos2021} through coupled cluster calculations. The molecular geometries were obtained in ref \citen{Loos2021} by optimizing the vacuum ground state structure at the CCSD(T) or CC3 level. Additionally, OO and TDDFT calculations are performed for intermolecular charge transfer excitations in two molecular dimers, tetrafluoroethene-ethene and ammonia-fluorine, at different intermolecular distances. The geometries of the tetrafluoroethene-ethene dimer are generated by varying the distance between the two monomers while keeping their internal structures fixed, starting from the dimer optimized in the ground state in vacuum using CC2 in ref \citen{Kozma2020}, while the geometries of the ammonia-fluorine dimer are obtained from a dimer optimized in the ground state in vacuum in ref \citen{Zhao2006} using a multi-configurational quadratic configuration interaction approach, MC-QCISD/3. The geometries of the two dimers correspond to those used in ref \citen{Bogo2024}, for which reference values of excitation energy and charge transfer distance calculated using equation-of-motion coupled cluster (EOM-CC) at the CCSD(T) level are available.

All calculations use the GGA functional PBE\cite{Perdew1996} and are spin-unrestricted. For the OO calculations, the frozen core approximation and the projector augmented wave (PAW) formalism\cite{Blochl1994} are used. The valence electrons for the systems with intramolecular and intermolecular charge transfer excitations are represented by an LCAO orbital basis set consisting of primitive Gaussian functions from the aug-cc-pVDZ\cite{Dunning1989, Kendall1992, Woon1994} and the cc-pVDZ sets, respectively. Each basis set is augmented with a single set of numeric atomic orbitals (referred to as Gaussian basis set + sz)\cite{Rossi2015, Larsen2009}. The TDDFT calculations of intermolecular charge transfer excited states use the cc-pVDZ basis set for all electrons. For the calculations on intermolecular charge transfer states, diffuse functions are not included, as was done for the EOM-CCSD(T) reference results of ref \citen{Bogo2024}. 

The OO calculations are performed with three different methods: (1) the FR-DO approach described in section \ref{sec:freeze_and_release}, (2) direct optimization in combination with MOM without constrained optimization step (DO-MOM), and (3) a conventional DIIS scheme together with MOM (SCF-MOM). For the FR-DO calculations, an L-BFGS algorithm\cite{Ivanov2021cpc} with a maximum step size of 0.2 is used for the constrained optimization step and a limited-memory symmetric rank 1 (L-SR1) algorithm\cite{Levi2020b} with a maximum step size of 0.1 is used for the full optimization after releasing the constraints. The DO-MOM calculations use L-SR1 with a maximum step size of 0.2, the default in GPAW designed to give a good balance of robustness and efficiency. The SCF-MOM calculations are based on direct diagonalization of the KS Hamiltonian matrix with Pulay mixing of the electron density\cite{Mortensen2024, Pulay1980}. Unless otherwise stated, the OO calculations are considered converged if a precision of 4$\cdot 10^{-8}$ eV$^2$ per valence electron in the squared residual of the KS equations,
\added[id=YLAS, comment = {R1C6}]{
\begin{equation}
    \frac{1}{N}\sum_{i = 1}^{M}\int d\matr{r}f_{i}\left|\matr{\hat{h}}_{\mathrm{KS}}\psi_{i}\left(\matr{r}\right) - \sum_{j = 1}^{M}\lambda_{ij}\psi_{j}\left(\matr{r}\right)\right|^{2}\,,
        \label{eq:residual}
\end{equation}
}
is achieved within 333 iterations (default in GPAW). In the initial, constrained optimization step of the FR-DO calculations, the orbitals are optimized to a less stringent precision of 4$\cdot 10^{-3}$ eV$^2$ per valence electron. \added[id=GL, comment=R1C4]{In the DO-MOM and FR-DO calculations, after convergence to the optimal orbitals, the Hamiltonian matrix is diagonalized within subspaces of equally occupied orbitals. For the FR-DO calculations, subspace diagonalization is also performed after the constrained optimization step. The molecular orbital energies are the eigenvalues resulting from subspace diagonalization.}

In the DO-MOM and SCF-MOM calculations, the MOM algorithm of ref\ \citenum{Barca2018} is employed, where at each iteration the occupation numbers are chosen such that the occupied orbitals are those with the largest projections into the occupied space of the initial guess orbitals
\begin{equation}
    \centering
    \omega_{i} = \sqrt{ \sum^N_{r=1} \left| 
\Omega_{ri} \right|^2 },
    \label{eq:mom} 
\end{equation}
with $N$ being the number of occupied orbitals, and $\Omega_{ri}$ being the overlap between occupied orbital $r$ of the initial guess and orbital $i$ at the current iteration,
\begin{equation}
\Omega_{ri} = \int \psi^{0*}_r(\vect{r}) \psi_i(\vect{r}) \mathrm{d} \vect{r}\,.
\end{equation}

All excited states considered in the present work are open-shell singlets. \replaced[id=GL, comment=R1C1]{However, the OO unrestricted KS solutions are spin-mixed with a spin contamination of $\sim$1 electrons, quantified as the integral of the negative part of the spin density, $\rho_\mr{s}(\mathbf{r}) = \rho_{\uparrow}(\mathbf{r}) - \rho_{\downarrow}(\mathbf{r})$: 
\begin{equation}
c = \int_{\rho_\mr{s}<0} -\rho_\mr{s}(\mathbf{r})\, \mr{d}\mathbf{r}
\end{equation}
}{, thus the OO unrestricted KS calculations provide spin-mixed solutions.} In all calculations presented here, no spin purification of the energy\,\cite{Ziegler1977} is applied. \replaced[id=GL, comment=R1C1]{For the intramolecular charge transfer states, the focus is on assessing the performance of the orbital optimization algorithms (for a comparison of the excitation energy calculated with local and semilocal functionals to higher-level references, see ref \citen{Selenius2024})}{ as the focus is on assessing the performance of the orbital optimization algorithms (for a benchmark of the performance of local and semilocal functionals with respect to intramolecular charge transfer excitations, see ref \citen{Selenius2024})}. For the intermolecular charge transfer states, for which a comparison with reference coupled cluster calculations is carried out, the effect of spin mixing on the energy \replaced[id=GL, comment=R1C1]{is found to be negligible: For the ammonia-fluorine dimer at the shortest intermolecular distance (3.5 Å), the difference between the spin-mixed energy and the spin-purified energy calculated according to $E_{\mr s} = 2E_{\mr m} - E_{\mr t}$, where $E_{\mr m}$ and $E_{\mr t}$ are the energy of the spin-mixed and triplet solutions, respectively, is only 3.7 meV with the PBE functional. At longer distances, the singlet-triplet gap is expected to decrease further, as shown by Bogo and Stein\cite{Bogo2024}. Therefore, spin purification is not needed.}{ has been shown to be negligible\cite{Bogo2024} since spin-mixed and triplet solutions become degenerate for long distances between donor and acceptor.} 

In addition to the value of excitation energy, OO solutions are characterized by computing a charge transfer distance according to the metric introduced by Le Bahers et al.\cite{LeBahers2011}
\begin{equation}
    \centering
    d^{\mathrm{CT}} = \dfrac{\left| \int \Delta \rho(\boldsymbol{r}) \boldsymbol{r} \, \mr{d}\boldsymbol{r} \right|}{q^{\mathrm{CT}}}\,,
    \label{eq:dct} 
\end{equation}
where $\Delta \rho(\boldsymbol{r})$ is the electron density difference between the excited state and the ground state, and $q^{\mathrm{CT}}$ is the charge transferred in the excitation evaluated as the integral of the positive part of $\Delta \rho(\boldsymbol{r})$. \added[id=GL, comment=R1C1]{For the intermolecular charge transfer excitations, as for the energy, also the charge transfer distance is found to be negligibly affected by spin purification: For the ammonia-fluorine dimer at the shortest intermolecular distance, the charge transfer difference computed from the spin-mixed and spin-purified PBE densities differ by only 0.007 Å. Therefore, the reported OO-calculated charge transfer distance is the one obtained for the spin-mixed solution.} For the TDDFT calculations, $\Delta \rho(\boldsymbol{r})$ is obtained from the relaxed density matrix computed using the Z-vector approach\cite{Pastore2017, Furche2002, Handy1984}.

The OO calculations are carried out with a development version\cite{GPAWDevBranch} of the GPAW software\cite{Mortensen2024, Enkovaara2010, Mortensen2005}, where the FR-DO method is implemented. For the TDDFT calculations, version 5.0.4 of the ORCA software\cite{Neese2022, Neese2012} is used. 

\section{Results}
\subsection{Intramolecular charge transfer states}
Table \ref{tab:iterations} reports the number of convergence failures and variational collapses as well as the average and maximum number of iterations for FR-DO, DO-MOM, and SCF-MOM calculations of the intramolecular charge transfer excited states of organic molecules in the benchmark set of Loos et al.\cite{Loos2021}. The number of iterations until convergence is also reported in Figure\ S1 of the Supporting Information (SI). \added[id=YLAS, comment={R2C2}]{For FD-DO, the iteration count is the sum of the iterations of both constrained and unconstrained optimization steps.}
\begin{table}[hbtp]
\scriptsize
  \caption{Number of convergence failures, variational collapses, average, and maximum number of iterations of FR-DO, DO-MOM, and SCF-MOM calculations of a set of 27 intramolecular charge transfer states of 15 organic molecules\cite{Loos2021} using the PBE functional and an aug-cc-pVDZ+sz basis set.}
  \label{tab:iterations}
  \begin{tabular}{|l|c|c|c|}
    \hline
    & FR-DO & DO-MOM & SCF-MOM\\
    \hline
    Convergence failures & 0 & 0 & 2\\
    Variational collapses & 0 & 4 & 0\\
    Avg. no. of iterations\textsuperscript{a} & 17.6 & 11.7 & 19.2\\
    Max. no. of iterations\textsuperscript{a} & 55 & 18 & 139\\
    Avg. no. of iterations\textsuperscript{b} & 17.6 & 12.8 & 19.2\\
    Max. no. of iterations\textsuperscript{b} & 55 & 25 & 139\\
    \hline
  \end{tabular}\\
  \vspace{5pt}
  \raggedright
\textsuperscript{a} Excluding unconverged and variationally collapsed calculations\\
\textsuperscript{b} Excluding unconverged calculations
\end{table}
\normalsize
The target excited state solutions have been identified in a previous work \cite{Selenius2024}. A calculation is considered collapsed if it converges to a solution with both excitation energy and charge transfer distance lower than those of the target solution, resulting in a greater deviation from the theoretical best estimate values. Table S1 in the SI reports all computed values of excitation energy and charge transfer distance, along with the corresponding theoretical best estimates.

SCF-MOM shows no variational collapses, but fails to converge in two cases. This behavior aligns with previous studies showing that DIIS-based MOM algorithms can suffer from erratic convergence, particularly for charge-transfer excitations\cite{Ivanov2021, Hait2020, Levi2020a, Levi2020b}. Excluding the unconverged cases, SCF-MOM requires an average of $\sim$19 iterations to reach convergence. In contrast, DO-MOM converges in all cases but exhibits four variational collapses. When the calculations that collapsed are excluded, DO-MOM converges in an average of $\sim$12 iterations. FR-DO converges to the target excited state solution in all cases, without variational collapse. While its average iteration count is higher than that of DO-MOM by approximately six iterations, it still outperforms SCF-MOM in terms of efficiency. As shown in Figure S1, the most challenging cases across all methods are the excited states of the $N$-Phenylpyrrole (PP) and benzonitrile molecules. In these instances, DO-MOM often converges faster but at the cost of a higher likelihood of variational collapse, whereas FR-DO consistently converges to the target solution.

In the following sections, the convergence properties of DO-MOM and FR-DO are analyzed in more detail by comparing the saddle point order of the target solution estimated before and after constrained optimization, for the entire set of intramolecular charge transfer states. Then, using a representative case, the A$_1$ charge transfer excited state of the PP molecule in its $90^{\circ}$-twisted conformation, it is demonstrated how MOM is unable to prevent variational collapse while FR provides a route to robustly converge on the target saddle point.

\subsubsection{Estimated saddle point order} \label{sec:spo}
\begin{figure}
\centering
\includegraphics[width = 1\textwidth]{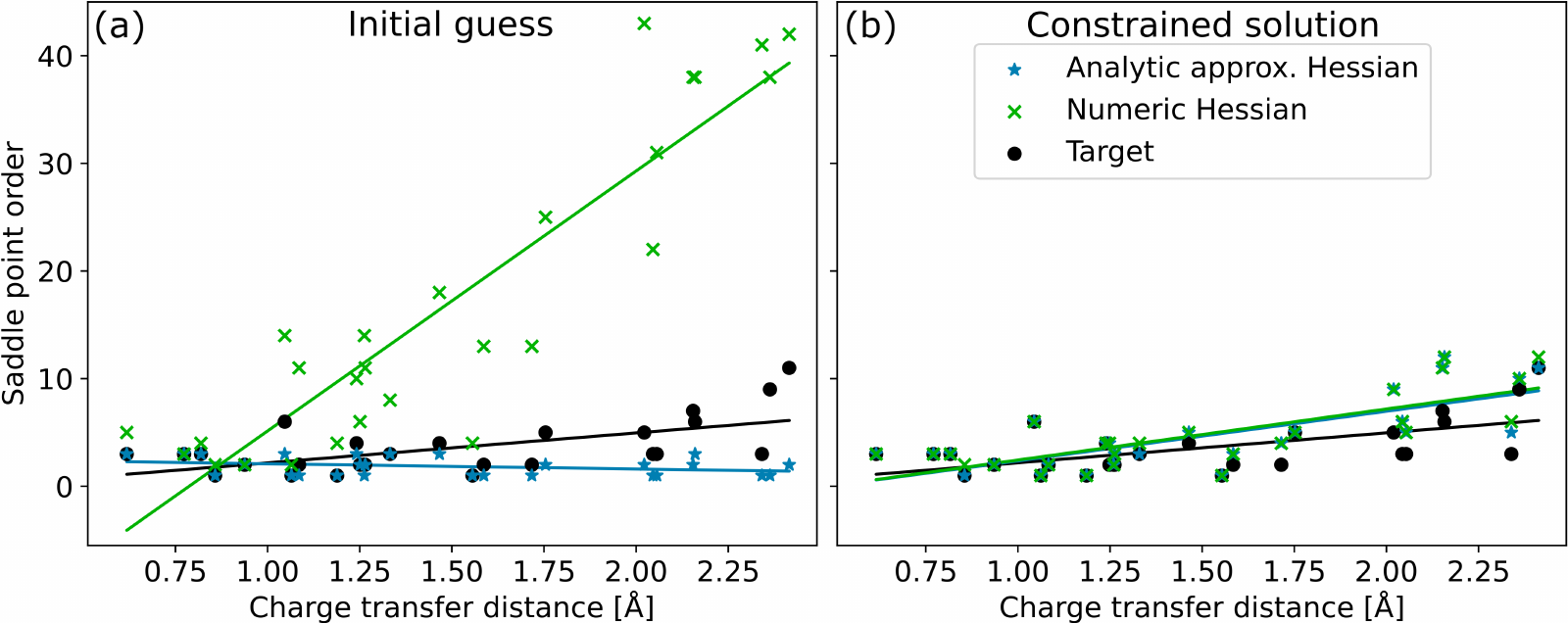}
\caption{Estimated (blue stars and green crosses) and true (black circles) saddle point order of the target solution for the set of intramolecular charge transfer excited states calculated with PBE/aug-cc-pVDZ+sz as a function of the charge transfer distance of the target solution. The saddle point order is obtained from the number of negative elements of the analytic diagonal electronic Hessian approximation (blue stars) and the number of negative eigenvalues of the numeric Hessian (green crosses). The lines represent linear regressions. At the initial guess of ground state orbitals (a), the analytic approximation tends to underestimate the saddle point order for large charge transfer distance, while the numeric Hessian considerably overestimates it. At the constrained solution obtained after constrained optimization (b) the saddle point estimate is significantly improved (average deviation of 0.7 for the numeric Hessian, if small negative eigenvalues are excluded, see Table\ S2).}
\label{fig:spos}
\end{figure}

Figure \ref{fig:spos} compares the saddle point order estimated at the initial guess made of ground state orbitals with changed occupations with that of the partially relaxed solution after constrained optimization for all intramolecular charge transfer excited states from Loos' benchmark set\cite{Loos2021}. The saddle point order of the final, target solutions obtained in the FR-DO calculations is also shown.
The saddle point order is obtained as the number of negative eigenvalues of the electronic Hessian evaluated numerically using a Davidson algorithm\cite{Schmerwitz2023}, and as the number of negative elements of the analytic diagonal approximation to the Hessian (see eq \ref{eq:precond}), in the case of the initial guesses. The saddle point order of the initial guesses and final solutions is also reported in Table S2 of the SI. 

At the initial guess made of ground state orbitals, the numeric and approximate analytic Hessians provide good estimates of the saddle point order for states with small charge transfer distance. However, for intermediate and large charge transfer distances, they significantly overestimate and underestimate the saddle point order, respectively. For the A$_1$ state of PP, the state with biggest charge transfer distance, the numeric Hessian gives 31 too many negative eigenvalues, while the preconditioner gives 9 too few negative eigenvalues, compared to the solution of the target excited state. The constrained optimization leads to a considerable improvement in the estimated  saddle point order, with the numeric and approximate analytic Hessians being in agreement with each other. For states with large charge transfer distance, the numeric and approximate analytic Hessian tend to slightly overestimate the number of negative eigenvalues. However, in those cases the magnitude of some of the negative eigenvalues is small. Table S2 reports the saddle point order estimated at the constrained solution by considering only negative eigenvalues with absolute value bigger than 1\,eV. In most cases, excluding the small eigenvalues gives a saddle point order in better agreement with the saddle point order of the final solution.

Therefore, it appears that the constrained optimization provides a set of partially relaxed excited state orbitals from which the degrees of freedom of negative curvature of the target solution can be more accurately estimated. This information can then be used to steer a subsequent unconstrained optimization to the target saddle point, without variational collapse, as illustrated in detail below for the A$_1$ charge transfer excited state of the twisted PP molecule.

\subsubsection{Twisted $N$-Phenylpyrrole}\label{sec:PP}
\begin{figure}[hbt]
    \centering
    \includegraphics[width=0.45\textwidth]{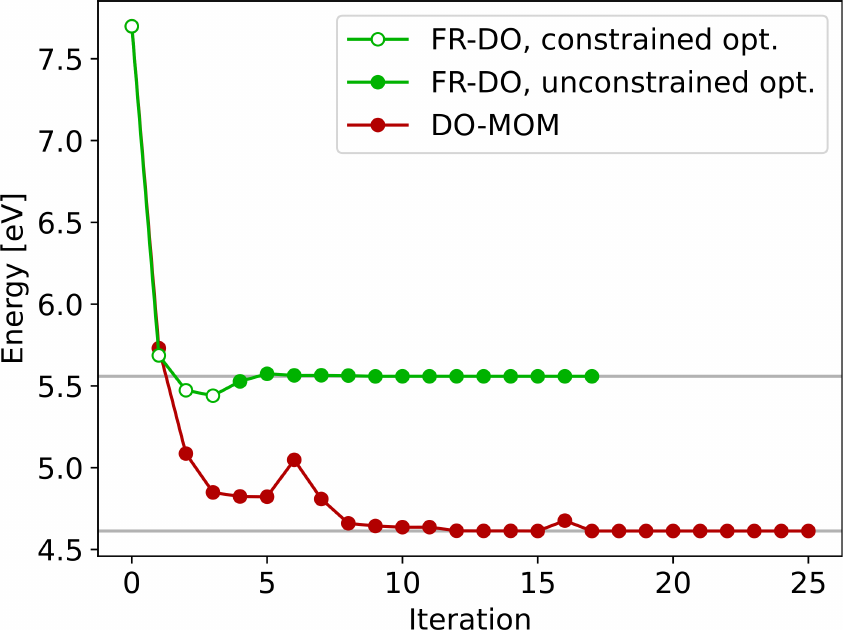}
    \caption{Convergence of the excitation energy in DO-MOM and FR-DO calculations of the spin-mixed A$_1$ charge transfer excited state of twisted $N$-Phenylpyrrole using PBE and an aug-cc-pVDZ+sz basis set. FR-DO converges to a charge-localized solution with excitation energy of 5.56\,eV, close to the theoretical best estimate (5.65\,eV), while DO-MOM collapses to a lower-energy (4.61\,eV), charge-delocalized solution (see also Figure \ref{fig:orb}).}
   \label{fig:conv}
\end{figure}

Figure \ref{fig:conv} shows the convergence of the energy in DO-MOM and FR-DO calculations of the A$_1$ charge transfer excited state of the twisted PP molecule.
The DO-MOM calculation converges after 24 iterations to a solution with an excitation energy of 4.61\,eV, while FR-DO converges in 16 iterations to a higher-energy solution with an excitation energy of 5.56\,eV. The latter is significantly closer to the theoretical best estimate excitation energy of 5.65\,eV obtained in CCSDT calculations\cite{Loos2021}. Moreover, the FR-DO solution is characterized by a significantly larger dipole moment and charge transfer distance (9.36 D and 2.41 Å, respectively) compared to the DO-MOM solution (3.33 D and 2.06 Å, respectively). The character of the FR-DO solution is in better agreement with CAM-B3LYP TDDFT calculations using the same basis set, which have previously been shown to perform well for the charge transfer excited states of PP and yield a dipole moment and charge transfer distance of 10.16 D and 2.41 Å, respectively\cite{Selenius2024}. 

\begin{figure}[h!]
    \centering
    \includegraphics[width=0.75\textwidth]{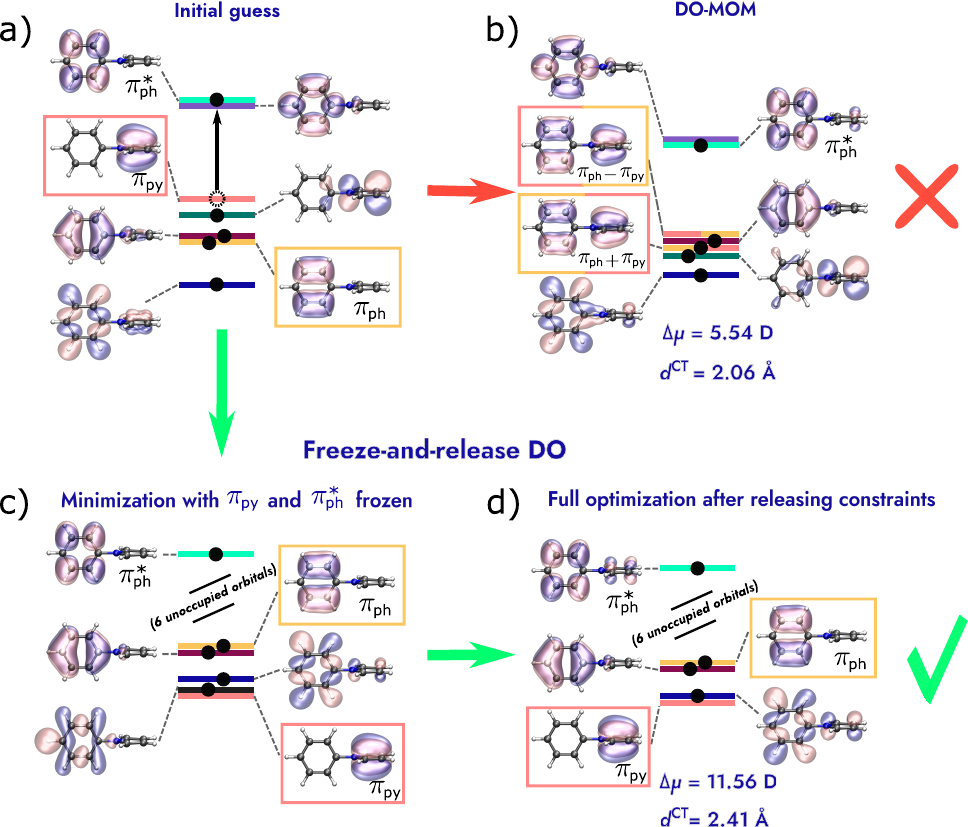}
    \caption{Orbitals of the initial guess, DO-MOM and FR-DO solutions for the calculation of the spin-mixed A$_1$ charge transfer excited state of the twisted $N$-Phenylpyrrole molecule using PBE/aug-cc-pVDZ+sz. (a) The initial guess consists of ground state orbitals with occupations changed according to a $\pi^*_{\mr{ph}} \leftarrow \pi_{\mr{py}}$ excitation (HOMO to LUMO+1). (b) DO-MOM converges to a 2\textsuperscript{nd}-order saddle point with small change in the dipole moment compared to the ground state and small charge transfer distance, where the $\pi_{\mr{py}}$ hole and a $\pi_{\mr{ph}}$ occupied orbital are mixed by $\sim$45$^\circ$. (c) The first step of constrained optimization in FR-DO prevents $\pi_{\mr{py}}$ and $\pi_{\mr{ph}}$ from mixing. (d) When the constraints are released, the calculation converges to a 10\textsuperscript{th}-order saddle point with larger charge transfer distance and change in dipole moment, in agreement with the results of  CAM-B3LYP TDDFT calculations (2.41 Å and 12.37 D) \cite{Selenius2024}. The opposite direction of the dipole moments in the ground and excited state are taken into account in the differences. The orbitals are visualized for isosurface values of $\pm 0.08$\,Å$^{-3}$. Some of the orbitals are omitted for clarity.
    }
   \label{fig:orb}
\end{figure}

Figure \ref{fig:orb} shows the initial guess orbitals and the orbitals of the DO-MOM and FR-DO solutions for the calculation of the A$_1$ excited state of twisted PP. The calculations are started by exciting an electron from the HOMO of a ground state calculation, which is largely localized on the pyrrole group ($\pi_{\mr{py}}$), to the ground state LUMO+1, primarily localized on the phenyl group ($\pi^*_{\mr{ph}}$).
In the initial guess, the electron hole created by the excitation is localized on the pyrrole group ($\pi_{\mr{py}}$ orbital). However, DO-MOM collapses to a solution with a hole delocalized over the entire molecule and nearly degenerate with a similarly delocalized occupied orbital, leading to a reduced charge transfer. The pair of delocalized occupied-unoccupied orbitals arise from mixing between the unoccupied $\pi_{\mr{py}}$ orbital and a lower-energy occupied orbital localized on the phenyl group, $\pi_{\mr{ph}}$. FR-DO instead avoids collapse along $\kappa_{\pi_{\mr{ph}} \pi_{\mr{py}}}$, giving a solution where $\pi_{\mr{ph}}$ and $\pi_{\mr{py}}$ are still localized on the phenyl and pyrrole groups, respectively.

\begin{figure}[h!]
    \centering
    \includegraphics[width=0.75\textwidth]{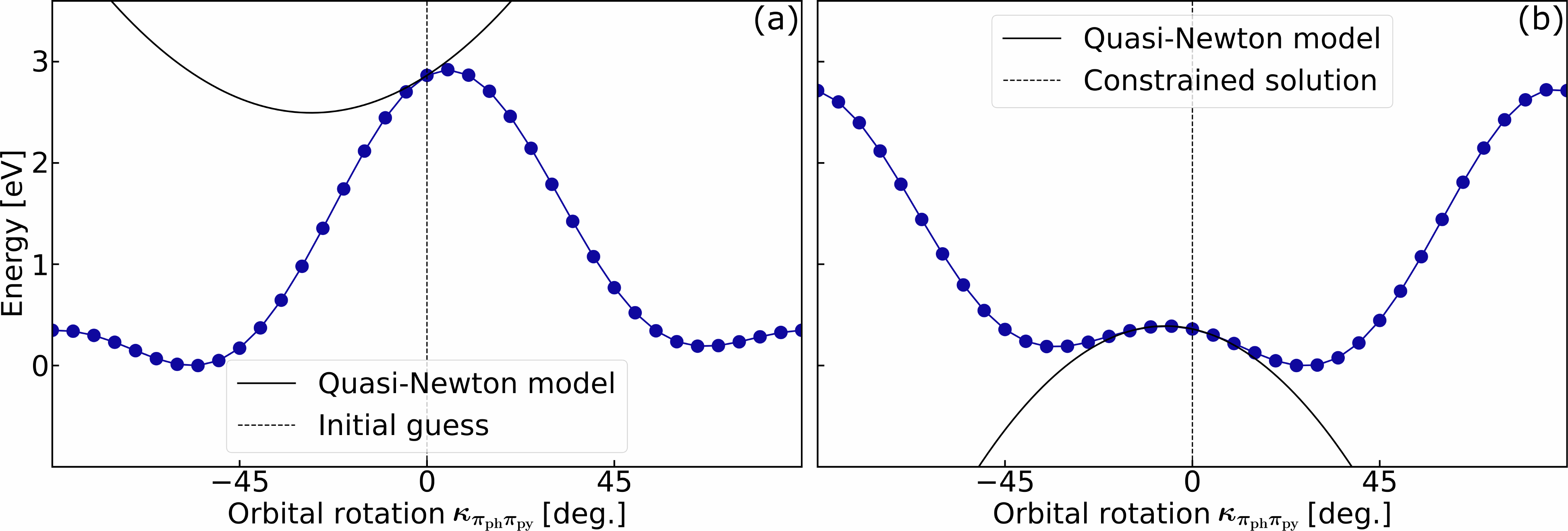}
    \caption{\replaced[id=GL]{Total electronic energy of twisted $N$-Phenylpyrrole (PP) in the A$_1$ charge transfer excited state, relative to the minimum,}{Energy of twisted $N$-Phenylpyrrole} as a function of the rotation angle $\kappa_{\pi_{\mr{ph}} \pi_{\mr{py}}}$ mixing the occupied $\pi_{\mr{ph}}$ and unoccupied $\pi_{\mr{py}}$ orbitals (see Figure \ref{fig:orb}) obtained with PBE/aug-cc-pVDZ+sz. \added[id=GL]{The energy is scanned starting from the initial guess ground state orbitals (a) and starting from the orbitals after the constrained optimization step of FR-DO (b).}  The minima of the energy along $\kappa_{\pi_{\mr{ph}} \pi_{\mr{py}}}$ correspond to spurious, charge-delocalized solutions, while the target charge-localized solution corresponds to the maximum close to 0$^{\circ}$. The black continuous curves represent the quasi-Newton model based on the energy gradient and Hessian approximation (eq \ref{eq:precond}) \replaced[id=GL]{at the ground state orbitals initial guess (a) and at the constrained solution (b). At the ground state orbitals initial guess, the quadratic model incorrectly gives a positive curvature, and DO-MOM takes a step towards a minimum.}{at the initial guess. In a DO-MOM calculation (a), the quadratic model incorrectly gives a positive curvature.} After constrained optimization in FR-DO (b), the model predicts the correct negative curvature and the quasi-Newton step is toward the saddle point.}
   \label{fig:scan}
\end{figure}

\begin{figure}[hbt]
    \centering
    \includegraphics[width=0.45\textwidth]{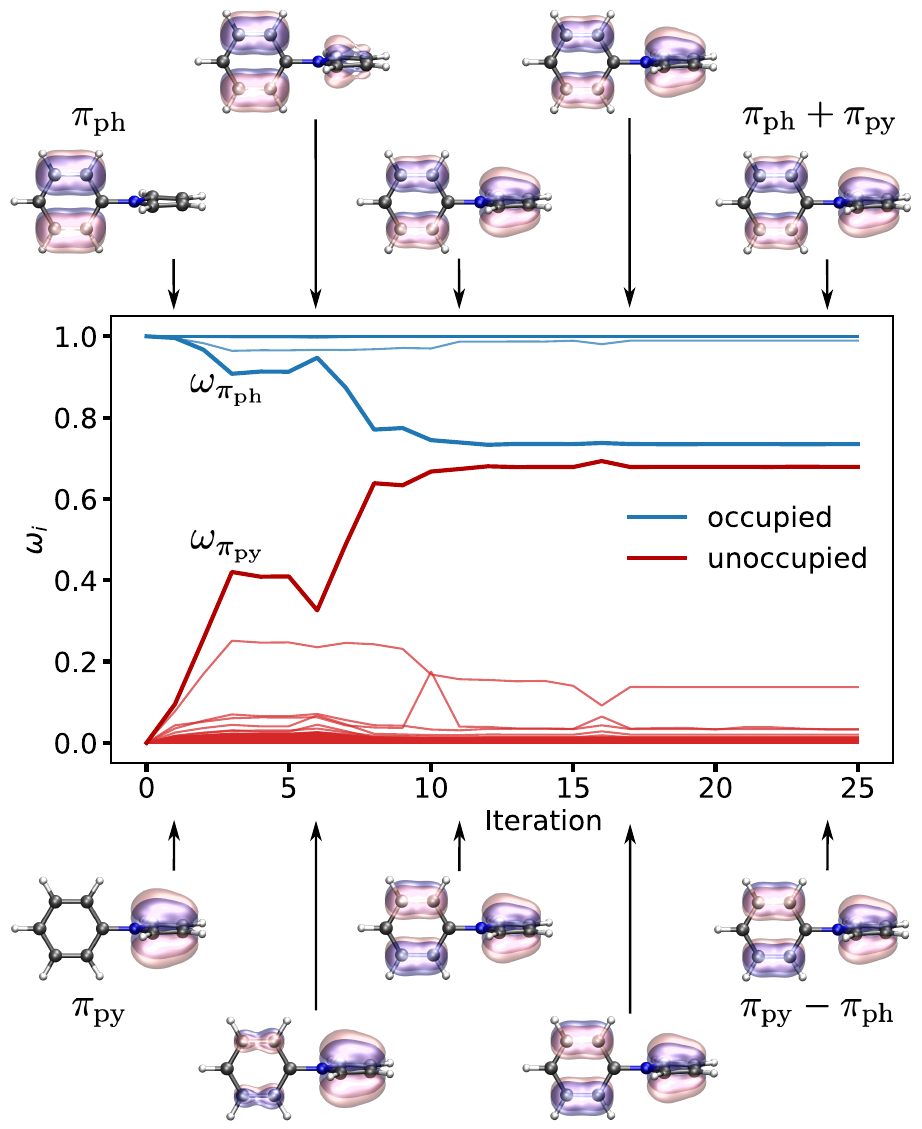}
    \caption{Orbital projections according to eq\ \ref{eq:mom} used to choose the occupation numbers in a DO-MOM calculation of the spin-mixed solution of the A$_1$ charge transfer excited state of the twisted $N$-Phenylpyrrole molecule with PBE/aug-cc-pVDZ+sz. $\omega_{\pi_{py}}$ and  $\omega_{\pi_{ph}}$ are the projections for the $\pi_{\mr{ph}}$ and $\pi_{\mr{py}}$ orbitals, respectively. These orbitals are visualized for selected iterations with isosurface values of $\pm 0.1$\,Å$^{-3}$, illustrating how they mix during the optimization.
    }
   \label{fig:mom}
\end{figure}

Figure \ref{fig:scan} shows a scan of the \replaced[id=GL, comment=R1C5]{total electronic energy, relative to the minimum,}{energy} along the degree of freedom corresponding to the rotation between the $\pi_{\mr{ph}}$ and $\pi_{\mr{py}}$ orbitals, $\kappa_{\pi_{\mr{ph}}\pi_{\mr{py}}}$\added[id=GL, comment=R1C5]{, in the A$_1$ excited state of twisted PP.} 
\replaced[id=GL, comment=R1C5]{The energy is scanned once starting from the initial guess ground state orbitals and once starting from the orbitals after the constrained optimization step of FR-DO, where the $\pi_{\mr{py}}$ and $\pi^*_{\mr{ph}}$ orbitals involved in the excitation are frozen. Along the $\kappa_{\pi_{\mr{ph}}\pi_{\mr{py}}}$}{Along this} orbital rotation, the energy has a maximum close to $0^{\circ}$ and a minimum close to $45^{\circ}$. The preconditioner evaluated as the inverse of the diagonal approximation to the electronic Hessian in eq \ref{eq:precond} at the initial guess of ground state orbitals has two negative elements, since there are two unoccupied orbitals below one occupied orbital at the initial guess (see Figure~\ref{fig:orb}). The component of the preconditioner along $\kappa_{\pi_{\mr{ph}} \pi_{\mr{py}}}$ is positive because the unoccupied orbital $\pi_{\mr{py}}$ has higher energy than the occupied orbital $\pi_{\mr{ph}}$. As a result, \added[id=GL, comment=R1C5]{the quadratic quasi-Newton model at the initial guess of ground state orbitals gives a positive curvature along $\kappa_{\pi_{\mr{ph}} \pi_{\mr{py}}}$, and} DO-MOM takes a step toward the direction of the negative of the gradient, \deleted[id=GL]{along $\kappa_{\pi_{\mr{ph}}\pi_{\mr{py}}}$,} leaving the concave region of the energy surface and going downhill toward the stationary point where the $\pi_{\mr{ph}}$ and $\pi_{\mr{py}}$ orbitals are mixed. This solution is a 2\textsuperscript{nd}-order saddle point, consistent with the number of negative elements of the preconditioner. Figure~\ref{fig:mom} shows the orbital projections computed according to eq~\ref{eq:mom} and used by MOM as weights to assign occupation numbers at each step of the wave function optimization.
Initially, the (occupied) $\pi_{\mr{ph}}$ and (unoccupied) $\pi_{\mr{py}}$ orbitals are localized on the phenyl and pyrrole parts of the molecule and their projections  according to eq\ \ref{eq:mom} are 1 and 0, respectively. As the calculation collapses to the charge-delocalized solution, the projections approach $1 / \sqrt{2} \approx 0.7$, corresponding to a mixing of $\sim45^\circ$. MOM is unable to prevent the variational collapse to a minimum along $\kappa_{\pi_{\mr{ph}} \pi_{\mr{py}}}$ as there are no orbitals with larger overlaps with the initially localized $\pi_{\mr{ph}}$ orbital among the manifold of unoccupied orbitals. 

In the constrained optimization step of FR-DO, the $\pi_{\mr{ph}}$ and $\pi_{\mr{py}}$ orbitals are frozen, so no steps are taken along $\kappa_{\pi_{\mr{ph}}\pi_{\mr{py}}}$ that would leave the concave region of the energy surface along this degree of freedom. The constrained optimization stabilizes the orbitals localized on the phenyl group and destabilizes the orbitals localized on the pyrrole group (see Figure \ref{fig:orb}). As a result, the preconditioner evaluated using the partially relaxed orbitals has 11 negative elements. One of the negative components of the preconditioner is along $\kappa_{\pi_{\mr{ph}}\pi_{\mr{py}}}$. Therefore, \added[id=GL, comment=R1C5]{the quadratic model predicts a negative curvature along $\kappa_{\pi_{\mr{ph}}\pi_{\mr{py}}}$, (see Figure \ref{fig:scan})} and when the constraints are released, a step is taken in the direction of the positive gradient, \deleted[id=GL, comment=R1C5]{along $\kappa_{\pi_{\mr{ph}}\pi_{\mr{py}}}$,} thereby converging on the saddle point corresponding to the target charge-localized solution. The saddle point order of this solution is 11, which is consistent with the number of negative elements of the diagonal approximate Hessian evaluated using the orbitals after constrained optimization.

\subsection{Intermolecular charge transfer states}
Recently, Bogo and Stein\cite{Bogo2025, Bogo2024} reported that the squared-gradient minimization and MOM-based approaches exhibit variational collapse in density functional calculations of intermolecular charge transfer excited states of the tetrafluoroethene-ethene and ammonia-fluorine dimers. Table \ref{tab:interCTEnergy} shows the values of excitation energy of the intermolecular charge transfer states of these two dimers calculated for different intermolecular distances at the PBE/cc-pVDZ level of theory using FR-DO and DO-MOM, together with the results of reference EOM-CCSD(T) calculations by Bogo and Stein\cite{Bogo2024}. The intermolecular distance is defined as the shortest distance between atoms of the two monomers. 
\begin{table}[!h]
\scriptsize
  \caption{Excitation energy values (in eV) for intermolecular charge transfer excited states of the tetrafluoroethene-ethene and ammonia-fluorine dimers obtained in PBE/cc-pVDZ calculations using DO-MOM and FR-DO compared to reference EOM-CCSD(T) results\cite{Bogo2024}.}
  \label{tab:interCTEnergy}
  \begin{tabular}{|l|c|c|c|c|}
    \hline
    Molecule & Intermolecular distance [\AA] & DO-MOM & FR-DO & EOM-CCSD(T)\\
    \hline
    Tetrafluoroethene-ethene & 3.5  & 5.05 & 8.10 & 8.19\\
    Tetrafluoroethene-ethene & 4.25 & 8.90 & 8.90 & 9.68\\
    Tetrafluoroethene-ethene & 5.00 & --   & 9.61 & 9.94\\
    Ammonia-Fluorine         & 3.5  & 6.81 & 8.33 & 8.34\\
    Ammonia-Fluorine         & 4.25 & --   & 8.77 & 8.82\\
    Ammonia-Fluorine         & 5.00 & --   & 9.07 & 9.16\\
    Ammonia-Fluorine         & 8.00 & 9.75 & 9.75 & 9.97\\
    Ammonia-Fluorine         & 10.0 & 9.95 & 9.95 & 10.26\\
    \hline
  \end{tabular}
\end{table}

\begin{figure}
\centering
\includegraphics[width = 0.75\textwidth]{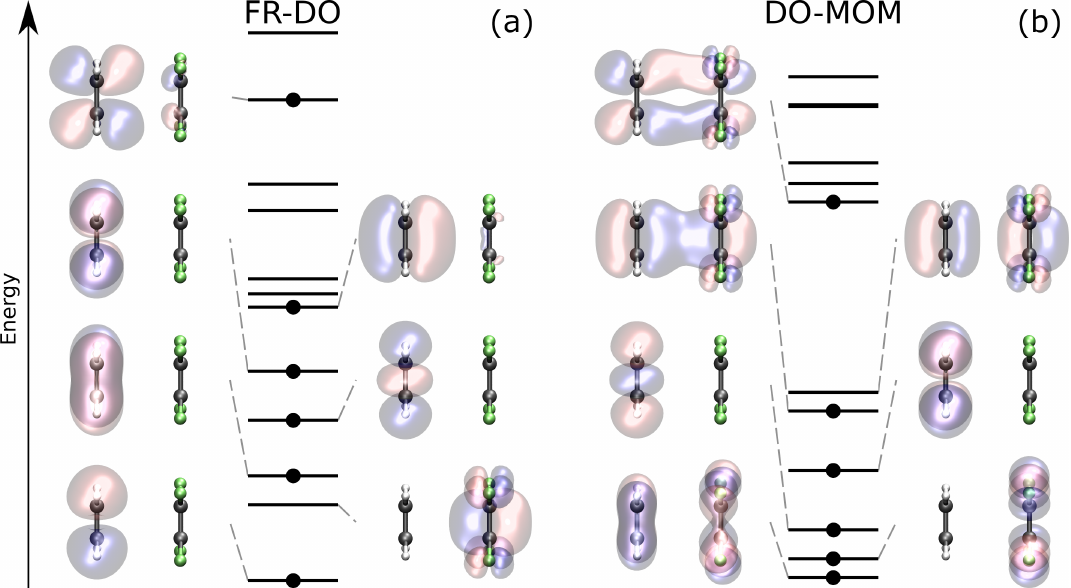}
\caption{Molecular orbitals of an intermolecular charge transfer excited state of the tetrafluoroethene-ethene dimer obtained with PBE/cc-pVDZ+sz for an intermolecular distance of 3.5\,\AA\ with (a) FR-DO and (b) DO-MOM. The FR-DO calculation converges to the desired charge transfer state, which corresponds to a 9\textsuperscript{th}-order saddle point and shows no significant mixing between orbitals localized on different fragments. The DO-MOM calculation collapses to a charge-delocalized solution, corresponding to a 1\textsuperscript{st}-order saddle point with mixing between orbitals localized on different fragments. The orbitals are visualized with an isosurface value of $\pm 0.08$\,Å$^{-3/2}$.}
\label{fig:MOsTFEE}
\end{figure}

The charge transfer excited state in the tetrafluoroethene-ethene dimer arises from excitation from the HOMO of the ethene molecule to the LUMO of the tetrafluoroethene molecule. For a dimer distance of 3.5\,\AA, DO-MOM shows a variational collapse leading to a charge-delocalized solution more than 3\,eV lower than the EOM-CCSD(T) result, while the solution obtained with FR-DO is charge localized and only 0.09 eV lower in energy than the EOM-CCSD(T) result. The molecular orbitals of the charge-delocalized solution obtained with DO-MOM and the solution with the expected charge transfer character obtained with FR-DO are illustrated in Figure \ref{fig:MOsTFEE}. 
While for FR-DO all orbitals are localized on either of the two molecular fragments, several orbitals are delocalized over both fragments in the solution obtained with DO-MOM, resulting in an artificial delocalization of the charge supposed to be transferred in the excitation. The FR-DO solution corresponds to a 9\textsuperscript{th}-order saddle point on the electronic energy surface, whereas the collapsed DO-MOM solution corresponds to a 1\textsuperscript{st}-order saddle point. At an intramolecular distance of 4.25\,\AA, both methods converge to the same charge-localized solution. DO-MOM fails to converge at an intramolecular distance of 5\,\AA, while FR-DO converges to a solution around 0.3 eV lower than EOM-CCSD(T).

\begin{figure}[h!]
\centering
\includegraphics[width = 1\textwidth]{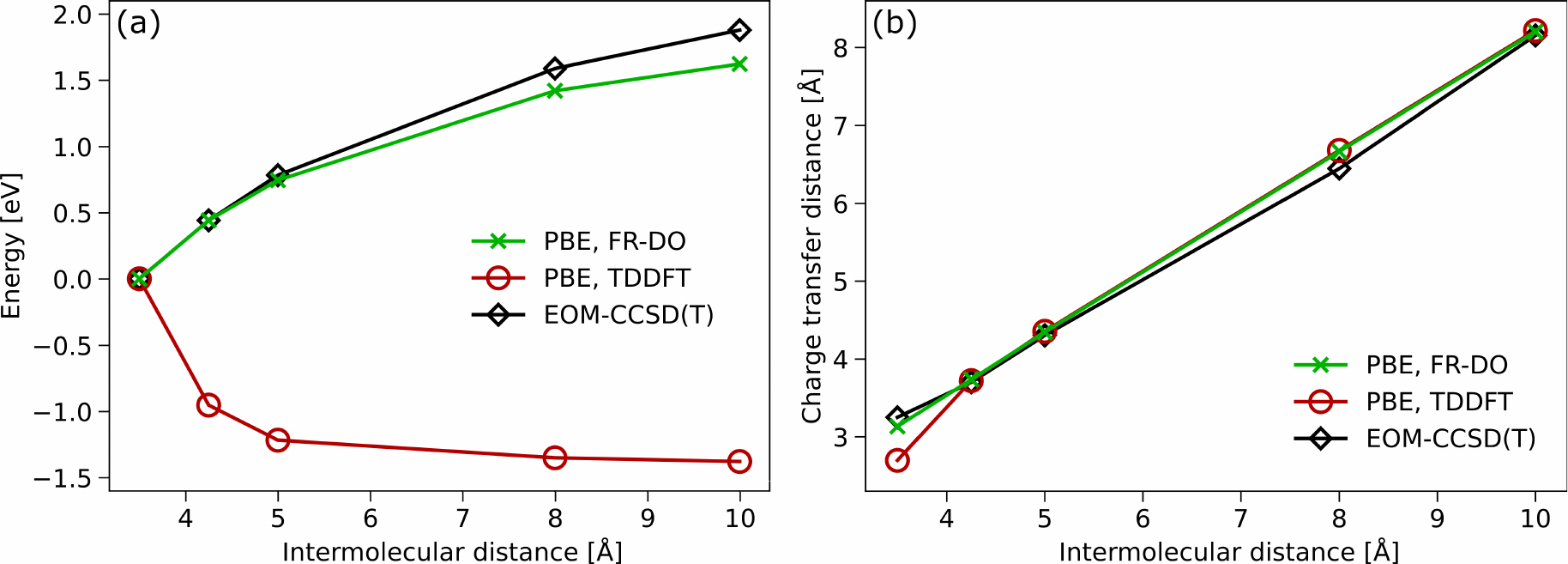}
\caption{(a) Change in the excitation energy and (b) the charge transfer distance of an intermolecular charge transfer excited state of the ammonia-fluorine dimer with respect to intermolecular distance from FR-DO (green crosses) and TDDFT (red circles) calculations using PBE/cc-pVDZ+sz, in comparison to reference EOM-CCSD(T) results\cite{Bogo2024} (black squares). The TDDFT energy curve shows a qualitatively incorrect trend, while the FR-DO provides the expected $1/R$ variation of the energy in close agreement with the EOM-CCSD(T) results.}
\label{fig:EandDCT}
\end{figure}

In the ammonia-fluorine dimer, the charge transfer excited state arises from excitation from the HOMO of the ammonia molecule to the LUMO of the fluorine molecule, as illustrated in Figure\ S3 of the SI. While DO-MOM shows a variational collapse to a charge-delocalized solution at a dimer distance of 3.5\,\AA\ and does not converge for distances of 4.25\,\AA\ and 5\,\AA, FR-DO systematically converges to a charge-localized solution (see Figure\ S3). Figure \ref{fig:EandDCT} shows the variation of the excitation energy and charge transfer distance of the excited state as a function of the intermolecular distance obtained with FR-DO and linear-response TDDFT with the PBE functional and the cc-pVDZ+sz basis set, as compared to a reference EOM-CCSD(T) excitation energy curve taken from ref\ \citenum{Bogo2024}. 
While TDDFT exhibits a qualitatively incorrect dependency of the energy as a function of the intermolecular distance, FR-DO agrees with the EOM-CCSD(T) quite well, showing the expected approximate $1/R$ dependency of the energy on the distance between donor and acceptor, even with a semilocal functional such as PBE. The charge transfer distances of all methods agree very well with the EOM-CCSD(T) reference values, with the exception of the TDDFT result for an intermolecular distance of 3.5\,\AA.

\section{Discussion and conclusions}
Intra- and intermolecular charge transfer excitations have so far remained challenging cases for variational excited state methods\cite{Selenius2024, Bogo2024, Mazzeo2023, Schmerwitz2023}. Here, we have shed light on some of the challenges that arise when using GGA functionals in orbital-optimized density functional calculations of charge transfer states. As shown for a large set of charge transfer excitations in organic molecules, the electronic Hessian at the commonly employed initial guess of ground state orbitals with nonaufbau occupation is characterized by a large number of negative eigenvalues, much higher than the saddle point order of the target solution, indicating that at such an initial guess, the energy surface is highly rugged. This likely leads to the convergence issues observed in calculations based on the DIIS SCF algorithm. 

Methods based on direct optimization of the orbitals are typically more robust than SCF approaches based on eigendecomposition of the Hamiltonian matrix. However, in the case of charge transfer excitations, the commonly used analytic diagonal approximation of the initial Hessian fails  to capture the structure of the electronic energy landscape, significantly underestimating the number of negative eigenvalues of the target solution, as demonstrated here. As a result, DO calculations of charge transfer states started from an initial guess made of ground state orbitals are prone to collapse to lower-energy solutions with too low saddle point order. It is found that the calculations collapse most often to charge-delocalized solutions where the hole and/or excited orbital are mixed with occupied or unoccupied orbitals. Such charge-delocalized solutions do not seem to correspond to physical states of the system, and are therefore considered spurious. Their presence on the electronic energy surface is found to affect calculations of both intra- and intermolecular charge transfer excitations.

Here, it is demonstrated that the maximum overlap method is unable by construction to prevent the variational collapse in such cases because the orbital rotations that produce the delocalized solutions typically involve mixing angles below 45$^{\circ}$, which is not detected as a variational collapse. The simple strategy proposed here to converge charge transfer excited states involves a first step of constrained optimization where the energy is minimized while the hole and excited electron orbital involved in the excitation are frozen. This moves the system out of the region of highly rugged energy surface, yielding an improved initial guess with a number of negative eigenvalues that is found to be close to the saddle point order of the target charge transfer solution. Then, the energy can be maximized along the estimated degrees of freedom of negative curvature in a subsequent step of direct unconstrained optimization converging on the target saddle point. This approach, which we refer to as freeze-and-release direct optimization (FR-DO), is shown to robustly converge PBE calculations of challenging intramolecular charge transfer states of organic molecules and intermolecular charge transfer states of molecular dimers. In the latter case, the energy obtained with the computationally efficient PBE functional varies approximately as $1/R$ with the separation between donor and acceptor for the intermolecular charge transfer excited states, in agreement with high-level coupled-cluster calculations. On the contrary, linear-response TDDFT with the same functional gives a qualitatively incorrect energy curve.

\added[id=GL, comment=R2C3]{The choice of the initial guess excitations to select the orbitals to freeze in the first step of constrained optimization is an important consideration. When targeting a specific excited state, the initial guess excitation(s) can be generated by a preliminary linear-response TDDFT calculation, similarly to the generation of the constraint potential in constrained orbital-optimized methods\cite{Kussmann2024}. If several excited states are wanted, for, e.g., the calculation of absorption spectra, the initial guess excitations can be generated as orbital rotations within an ``active space''. This also provides a way to include double excitations, which are missing in linear-response TDDFT calculations with the adiabatic approximation. As other single-determinant approaches, the FR-DO method remains limited to excitations dominated by a single configuration, and cannot adequately capture multiconfigurational states.}

While a quasi-Newton algorithm has been used here, the unconstrained optimization can also be carried out with a recently presented robust generalized mode following method\cite{Schmerwitz2023}, where the gradient along the degrees of freedom of negative curvature is inverted, recasting the saddle point optimization into a minimization. More recently, Bogo et al.\cite{Bogo2025} have employed the freeze-and-release approach with a squared-gradient minimization\cite{Hait2020} for the unconstrained optimization step. The same authors report that employing DIIS-based algorithms together with MOM in the unconstrained optimization step is still susceptible of variational collapse.

The present work has focused on PBE as a representative semilocal function. An important open question is how the inclusion of exact exchange modifies the structure of the energy landscape. \replaced[id=GL, comment=R1C8]{It will also be valuable to investigate how the number of multiple solutions and the saddle point order of excited states differ compared to state-specific orbital-optimized wave function approaches\cite{Marie2023, Burton2022, Kossoski2021}}{, the quality of the initial guesses, and the saddle point order of charge transfer excited states.} Our ongoing work aims to map these features in detail and assess the performance of FR-DO when applied to orbital-optimized calculations with global and range-separated hybrid functionals. \replaced[id=GL]{It is also important}{Finally, it will be valuable} to evaluate how self-interaction corrected\cite{Sigurdarson2023, Perdew1981} and recently developed state-specific\cite{Loos2025, Gould2025} and ensemble-based\cite{Gould2023} density functionals designed for excited states perform for both intra- and intermolecular charge transfer excitations. 

\added[id=GL, comment=R1C3]{Finally, the FR-DO strategy can be extended to excited state variational KS methods that incorporate spin purification, such as the restricted open-shell KS approach\cite{Frank1998, Filatov1998}, provided that the corresponding orbital gradients are available. It will then be interesting to examine the electronic energy landscape and the saddle point order of charge transfer excited state solutions within this framework. This extension will also make it possible to more straightforwardly perform geometry optimization and molecular dynamics simulations for open-shell singlet excited states.}

\begin{acknowledgement}
The authors thank Nicola Bogo and Christopher J. Stein for providing the reference values of excitation energy and charge transfer distance as well as geometries of the molecular dimers. The authors also thank Alec Elías Sigurdarson for useful discussions and comments on the manuscript. Y.L.A.S.\ acknowledges support by the Max Planck Society. E.S. and G.L acknowledge support by the Icelandic Research Fund (grants nos.\ 239678 and 2511544). G.L. acknowledges support from the ERC under the European Union's Horizon Europe research and innovation programme (grant no. 101166044, project NEXUS). Views and opinions expressed are however those of the author(s) only and do not necessarily reflect those of the European Union or ERC Executive Agency. Neither the European Union nor the granting authority can be held responsible for them. The authors acknowledge computer resources, data storage, and user support by the Icelandic Research e-Infrastructure (IREI), funded by the Icelandic Infrastructure Fund.

\end{acknowledgement}

\begin{suppinfo}

Values of excitation energy and charge transfer distance of the intramolecular charge transfer states obtained with FR-DO, DO-MOM, and SCF-MOM; estimated saddle point order at the initial guess and constrained solution, and saddle point order of the final solution for the intramolecular charge transfer states; number of iterations taken by FR-DO, DO-MOM, and SCF-MOM in the calculations of the intramolecular charge transfer states; excitation energy of the ammonia-fluorine dimer as a function of intermolecular distance computed with FR-DO, DO-MOM, and the theoretical best estimate curve; molecular orbitals of the initial guess, FR-DO solution, and DO-MOM solution for the charge transfer state of the ammonia-fluorine dimer at an internuclear distance of 3.5 Å.

Data related to the results presented in this article and instructions on the generation thereof are available at Zenodo.

\end{suppinfo}

\bibliography{main}

\end{document}


\begin{table}[hbtp]
\scriptsize
  \caption{Excitation energy (in eV) and charge transfer distance (in Å), $d^{\mathrm{CT}}$, of charge transfer states of organic molecules obtained from orbital optimized calculations using FR-DO, DO-MOM, and SCF-MOM with PBE/aug-cc-pVDZ+sz, together with the theoretical best estimate (TBE) values of excitation energy\cite{Loos2021}. }
  \label{tbl:energies}
  \begin{tabular}{|lc|c|rr|rr|rr|}
    \hline
    Molecule & Sym. & \multicolumn{1}{c}{TBE$\,^\textrm{a}$} & \multicolumn{2}{|c}{FR-DO} & \multicolumn{2}{|c}{DO-MOM} & \multicolumn{2}{|c|}{SCF-MOM}\\
     \hline
   & & \multicolumn{1}{c}{$\Delta$E} & \multicolumn{1}{|c}{$\Delta$E} & \multicolumn{1}{c}{$d^{\mr{CT}}$} & \multicolumn{1}{|c}{$\Delta$E} & \multicolumn{1}{c}{$d^{\mr{CT}}$} & \multicolumn{1}{|c}{$\Delta$E} & \multicolumn{1}{c|}{$d^{\mr{CT}}$}\\
    \hline
    Aminobenzonitrile (ABN)           & A$_1$ & 5.09 & 3.69 & 1.06 & 3.69 & 1.06 & 3.69 & 1.06\\
    Aniline                           & A$_1$ & 5.48 & 4.31 & 0.82 & 4.31 & 0.82 & 4.31 & 0.82\\
    Azulene                           & A$_1$ & 3.84 & 3.00 & 0.94 & 3.00 & 0.94 & 3.00 & 0.94\\
                                      & B$_2$ & 4.49 & 3.97 & 0.77 & 3.97 & 0.77 & 3.97 & 0.77\\
    Benzonitrile                      & A$_2$ & 7.05 & 6.57 & 1.04 & 5.96 & 0.58 & -    & -   \\
    Benzothiadiazole (BTD)            & B$_2$ & 4.28 & 3.08 & 1.19 & 3.08 & 1.19 & 3.08 & 1.19\\
    Dimethylaminobenzonitrile (DMABN) & A$_1$ & 4.86 & 3.53 & 1.55 & 3.53 & 1.56 & 3.54 & 1.56\\
    Twisted DMABN                     & A$_2$ & 4.12 & 3.56 & 2.04 & 3.56 & 2.04 & 3.56 & 2.05\\
                                      & B$_1$ & 4.75 & 4.21 & 1.75 & 4.21 & 1.75 & 4.21 & 1.75\\
    Dimethylaniline (DMAn)            & B$_2$ & 4.40 & 3.82 & 1.08 & 3.82 & 1.08 & 3.82 & 1.08\\
                                      & A$_1$ & 5.40 & 4.17 & 1.33 & 4.17 & 1.33 & 4.18 & 1.33\\
    Hydrogen Chloride                 & $\Pi$ & 7.88 & 7.33 & 0.86 & 7.33 & 0.86 & 7.33 & 0.86\\
    p-Nitroaniline                    & A$_1$ & 4.39 & 3.26 & 2.05 & 3.26 & 2.05 & 3.26 & 2.05\\
    Nitrobenzene                      & A$_1$ & 5.39 & 4.13 & 1.46 & 4.13 & 1.46 & 4.13 & 1.47\\
    Nitrodimethylaniline (NDMA)       & A$_1$ & 4.13 & 3.05 & 2.34 & 3.05 & 2.34 & 3.05 & 2.34\\
    Nitropyridine $N$-Oxide (NPNO)    & A$_1$ & 4.10 & 2.73 & 1.72 & 2.73 & 1.72 & 2.73 & 1.72\\
    $N$-Phenylpyrrole (PP)            & B$_2$ & 5.32 & 4.12 & 1.59 & 4.12 & 1.59 & -    & -   \\
                                      & A$_1$ & 5.86 & 5.14 & 2.02 & 4.69 & 1.85 & 5.14 & 2.03\\
    Twisted PP                        & B$_2$ & 5.58 & 5.26 & 2.36 & 4.58 & 2.04 & 5.27 & 1.49\\
                                      & A$_1$ & 5.65 & 5.56 & 2.41 & 4.61 & 2.06 & 5.56 & 2.41\\
                                      & A$_2$ & 5.95 & 5.40 & 2.15 & 5.40 & 2.15 & 5.40 & 2.15\\
                                      & B$_1$ & 6.17 & 5.42 & 2.16 & 5.42 & 2.16 & 5.42 & 2.16\\
    Phthalazine                       & A$_2$ & 3.91 & 3.10 & 1.26 & 3.10 & 1.26 & 3.11 & 1.26\\
                                      & B$_1$ & 4.31 & 3.45 & 1.26 & 3.45 & 1.26 & 3.45 & 1.26\\
    Quinoxaline                       & B$_2$ & 4.63 & 3.48 & 1.25 & 3.48 & 1.25 & 3.48 & 1.25\\
                                      & A$_1$ & 5.65 & 4.56 & 0.62 & 4.56 & 0.62 & 4.56 & 0.62\\
                                      & B$_1$ & 6.22 & 5.15 & 1.24 & 5.15 & 1.24 & 5.15 & 1.24\\
    \hline
  \end{tabular}\\
  \vspace{5pt}
  \raggedright
 $^\textrm{a}$ Theoretical best estimates obtained at the CCSDT/aug-cc-pVQZ level in ref.\ \citenum{Loos2021}
\end{table}

\begin{table}[!h]
\scriptsize
  \caption{Computed saddle point order of charge transfer excited states of organic molecules at the initial guess, after constrained optimization, and at the target solution. The calculations use the PBE functional and the aug-cc-pVDZ+sz basis set. The values in parentheses are the number of negative eigenvalues with an absolute value bigger than 1\,eV. Constrained optimization leads to a significant improvement in the estimated saddle point order for both a numeric eigendecomposition of the Hessian and a diagonal analytic approximation (preconditioner of eq~7 in the main text). 
  }
  \label{tbl:spo}
  \begin{tabular}{|p{4.8cm}p{0.5cm}|p{1.2cm}p{1.2cm}|p{1.2cm}p{1.2cm}|p{1.2cm}|}
    \hline
     Molecule & Sym. &  \multicolumn{2}{|c}{Initial guess} & \multicolumn{2}{|c|}{Constrained solution} & \multicolumn{1} {|c|}{Final solution} \\
     \hline
      & &   Num. Hessian & Precond. (eq~7) & Num. Hessian & Precond. (eq~7) & Num. Hessian\\
     \hline
    Aminobenzonitrile (ABN) & A$_1$ & 2 (1) & 1 & 1 (1) & 1 & 1\\
    Aniline & A$_1$ & 4 (2) & 3 & 3 (1) & 3 & 3\\
    Azulene & A$_1$ & 2 (2) & 2 & 2 (2) & 2 & 2\\
            & B$_2$ & 3 (3) & 3 & 3 (3) & 3 & 3\\
    Benzonitrile & A$_2$ & 14 (13) & 3 & 6 (6) & 6 & 6\\
    Benzothiadiazole (BTD) & B$_2$ & 4 (3) & 1 & 1 (1) & 1 & 1\\
    Dimethylaminobenzonitrile (DMABN) & A$_1$ & 4 (1) & 1 & 1 (1) & 1 & 1\\
    Twisted DMABN & A$_2$ & 22 (18) & 1 & 6 (3) & 6 & 3\\
                  & B$_1$ & 25 (22) & 2 & 5 (5) & 5 & 5\\
    Dimethylaniline (DMA) & B$_2$ & 11 (8) & 1 & 2 (1) & 2 & 2\\
                          & A$_1$ & 8 (4) & 3 & 4 (2) & 3 & 3\\
    Hydrogen Chloride & $\Pi$ & 2 (1) & 1 & 2 (1) & 1 & 1\\
    p-Nitroaniline & A$_1$ & 31 (19) & 1 & 5 (4) & 5 & 3\\
    Nitrobenzene & A$_1$ & 18 (14) & 3 & 5 (4) & 5 & 4\\
    Nitrodimethylaniline (NDMA) & A$_1$ & 41 (26) & 1 & 6 (4) & 5 & 3\\
    Nitropyridine $N$-Oxide (NPNO) & A$_1$ & 13 (9) & 1 & 4 (3) & 4 & 2\\
    $N$-Phenylpyrrole (PP) & B$_2$ & 13 (8) & 1 & 3 (3) & 3 & 2\\
                           & A$_1$ & 43 (36) & 2 & 9 (6) & 9 & 5\\
    Twisted PP & B$_2$ & 38 (31) & 1 & 10 (6) & 10 & 9\\
               & A$_1$ & 42 (35) & 2 & 12 (7) & 11 & 11\\
               & A$_2$ & 38 (29) & 2 & 11 (7) & 11 & 7\\
               & B$_1$ & 38 (34) & 3 & 12 (7) & 12 & 6\\
    Phthalazine & A$_2$ & 14 (12) & 1 & 2 (1) & 2 & 2\\
                & B$_1$ & 11 (8) & 2 & 3 (1) & 3 & 2\\
    Quinoxaline & B$_2$ & 6 (4) & 2 & 4 (2) & 4 & 2\\
                & A$_1$ & 5 (5) & 3 & 3 (3) & 3 & 3\\
                & B$_1$ & 10 (9) & 3 & 4 (4) & 4 & 4\\
    \hline
    \multicolumn{2}{|c|}{Avg. abs. deviation} & 13.6 (9.7) & 1.7 & 1.2 (0.7) & 1.1 & \\
    \multicolumn{2}{|c|}{Max. abs. deviation} & 38 (31) & 9 & 6 (4) & 6 & \\
    \hline
  \end{tabular} \\
\raggedright
\end{table}

\begin{figure}[!h]
\centering
\includegraphics[width = 1\textwidth]{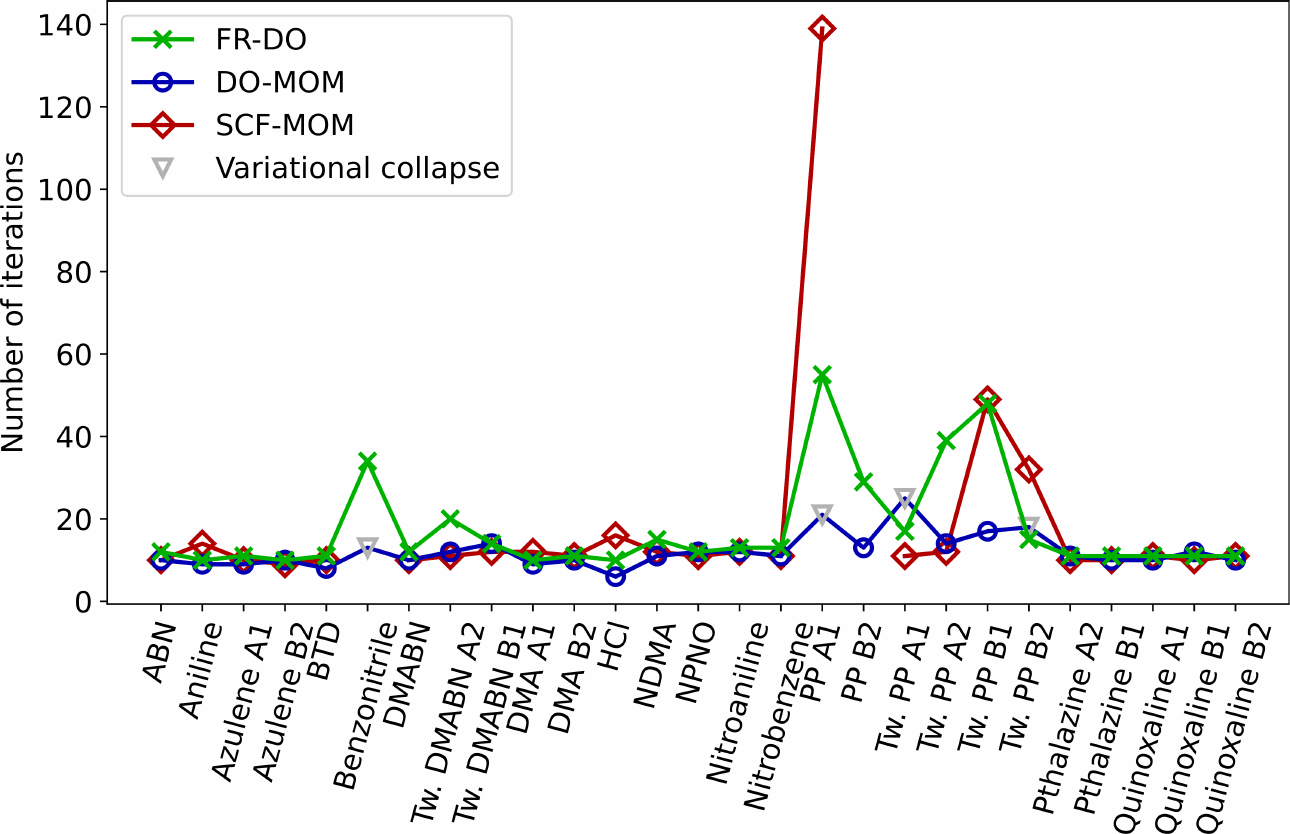}
\caption{Number of iterations taken by FR-DO (green crosses), DO-MOM (blue circles), and SCF-MOM (red diamonds) to converge each excited state in the set of intramolecular charge transfer states. Variational collapses are indicated by gray triangles.}
\label{fig:iterationsByState}
\end{figure}

\begin{figure}[h!]
    \centering
    \includegraphics[width=0.9\textwidth]{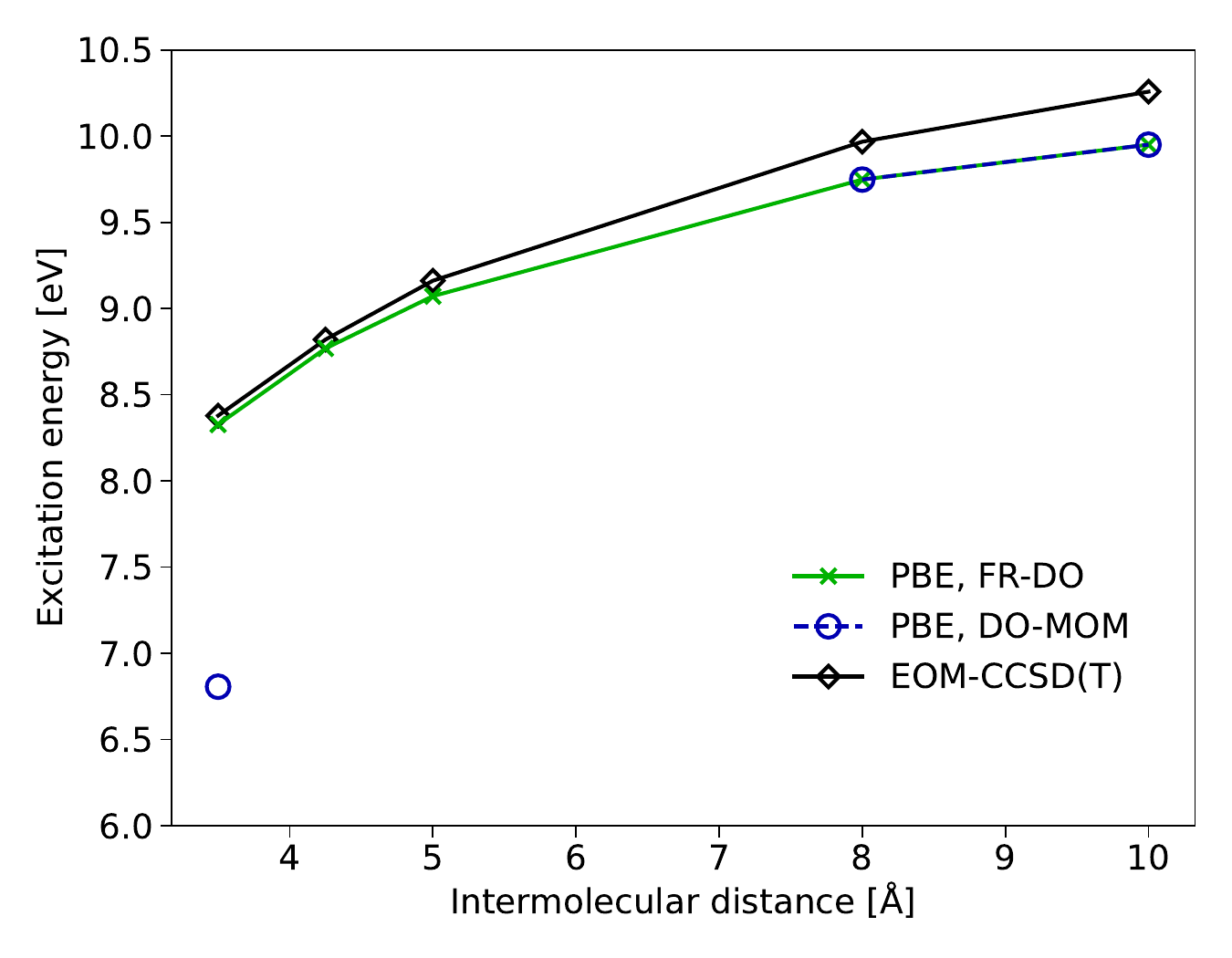}
    \caption{Excitation energy of the ammonia-fluorine dimer  as a function of the distance between the molecular fragments obtained in FR-DO (green crosses) and DO-MOM (blue circles) calculations with PBE/cc-pVDZ+sz, and taken from published EOM-CCSD(T) calculations\cite{Bogo2024} (black squares). The FR-DO energy curve agrees well with the many-body results. DO-MOM shows a variational collapse at an intermolecular distance of 3.5\,\AA\ and does not converge within 333 iterations until a distance of 8\,\AA\ is reached, above which it converges to the same solution as FR-DO.}
   \label{fig:ammonia}
\end{figure}

\begin{figure}[h!]
    \centering
    \includegraphics[width=1.0\textwidth]{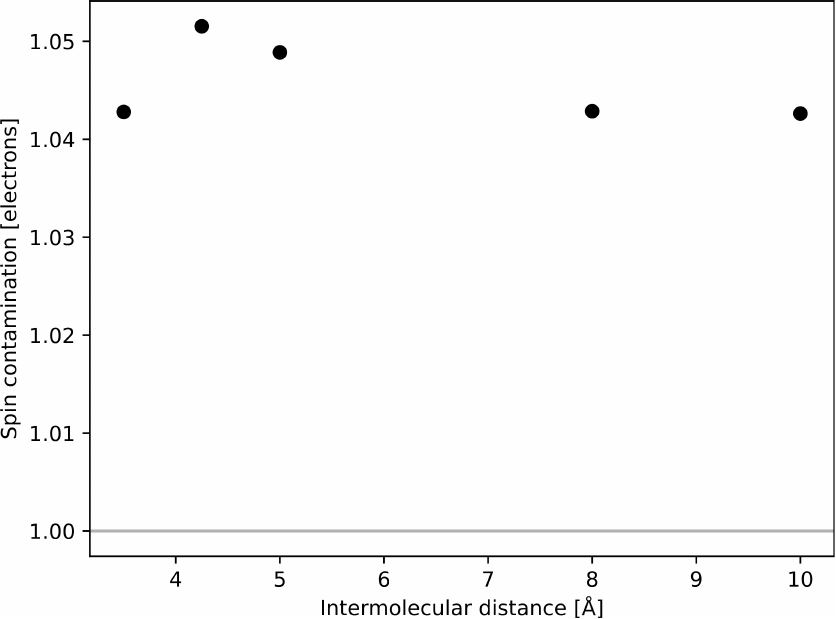}
    \caption{Spin contamination of an intermolecular charge transfer excited state of the ammonia-fluorine dimer with respect to intermolecular distance obtained in FR-DO calculations using PBE/cc-pVDZ+sz. The spin density is computed as the integral of the negative part of the spin density, $\rho_\mr{s}(\mathbf{r}) = \rho_{\uparrow}(\mathbf{r}) - \rho_{\downarrow}(\mathbf{r})$ (see eq 13 in the main text). A spin contamination value of 1 is highlighted by the gray horizontal line, indicating the value expected for a spin-mixed excited state solution.}
   \label{fig:ammonia}
\end{figure}

\begin{figure}[h!]
    \centering
    \includegraphics[width=1.0\textwidth]{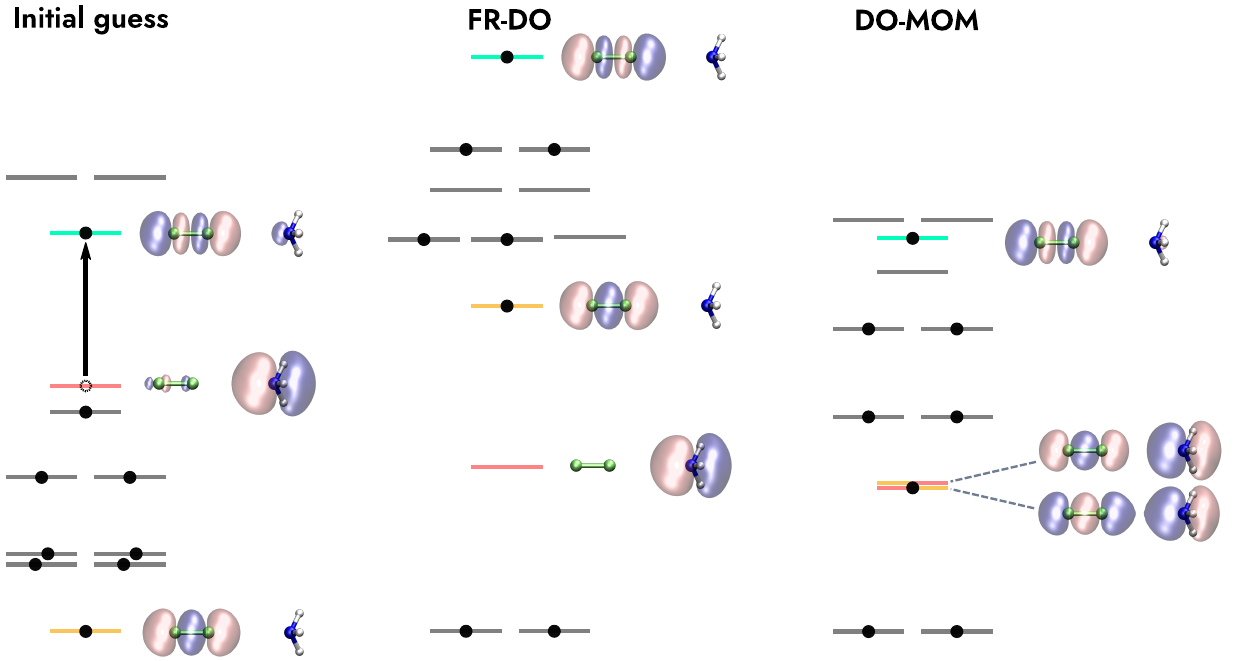}
    \caption{Molecular orbitals of the initial guess, the FR-DO solution, and the DO-MOM solution for the ammonia-fluorine dimer at an intermolacular distance of 3.5 Å. The orbitals are visualized with isosurface values of $\pm 0.08$\,Å$^{-3}$.}
   \label{fig:ammonia}
\end{figure}

\clearpage

\bibliography{si}